\documentclass[11pt, aps, preprint, nofootinbib,superscriptaddress,eqsecnum,titlepage]{revtex4-2}
\usepackage[active]{srcltx}
\usepackage[utf8]{inputenc}
\usepackage{enumerate}
\usepackage{bbold}
\usepackage{manfnt}
\usepackage{physics}
\usepackage{latexsym}
\usepackage{hyperref}
\usepackage{amsmath}
\usepackage{amsfonts}
\usepackage{graphicx}
\usepackage{slashed}
\usepackage{xcolor}
\usepackage{soul}
\usepackage{slashed}
\usepackage{braket}
\usepackage{natbib}
\usepackage{cancel}
\usepackage{MnSymbol}
\usepackage{wasysym}
\usepackage[export]{adjustbox}
\usepackage{graphicx}

\allowdisplaybreaks

\begin{document}
	
\title{\Large{{\sc Topological Superconductor from the Quantum Hall Phase: Effective Field Theory Description}}}

\author{M. Gomes}
\email{mgomes@fma.if.usp.br}
\affiliation{Instituto de F\'\i sica, Universidade de S\~ao Paulo\\
Caixa Postal 66318, 05315-970, S\~ao Paulo, SP, Brazil}
	
\author{Pedro R. S. Gomes}
\email{pedrogomes@uel.br}
\affiliation{Departamento de F\'\i sica, Universidade Estadual de Londrina,  86057-970, Londrina, PR, Brasil}

\author{K. Raimundo}
\email{kesley@usp.br }
\affiliation{Instituto de F\'\i sica, Universidade de S\~ao Paulo\\
Caixa Postal 66318, 05315-970, S\~ao Paulo, SP, Brazil}
	
\author{Rodrigo Corso B. Santos}
\email{rodrigocorso@uel.br}
\affiliation{Departamento de F\'\i sica, Universidade Estadual de Londrina,  86057-970, Londrina, PR, Brasil}
	
\author{A. J. da Silva}
\email{ajsilva@fma.if.usp.br}
\affiliation{Instituto de F\'\i sica, Universidade de S\~ao Paulo\\
Caixa Postal 66318, 05315-970, S\~ao Paulo, SP, Brazil}

	
\begin{abstract}

We derive low-energy effective field theories for the quantum anomalous Hall and topological superconducting phases. The quantum Hall phase is realized in terms of free fermions with nonrelativistic dispersion relation, possessing a global $U(1)$ symmetry. We couple this symmetry with a background gauge field and compute the effective action by integrating out the gapped fermions. In spite of the fact that the corresponding Dirac operator governing the dynamics of the original fermions is nonrelativistic, the leading contribution in the effective action is a usual Abelian $U(1)$ Chern-Simons term. The proximity to a conventional superconductor induces a pairing potential in the quantum Hall state, favoring the formation of Cooper pairs. When the pairing is strong enough, it drives the system to a topological superconducting phase, hosting Majorana fermions. Even though the continuum $U(1)$ symmetry is broken down to a $\mathbb{Z}_2$ one, we can forge fictitious $U(1)$ symmetries that enable us to derive the effective action for the topological superconducting phase, also given by a Chern-Simons theory. To eliminate spurious states coming from the artificial symmetry enlargement, we demand that the fields in the effective action are $O(2)$ instead of $U(1)$ gauge fields. In the $O(2)$ case we have to sum over the $\mathbb{Z}_2$ bundles in the partition function, which projects out the states that are not $\mathbb{Z}_2$ invariants. The corresponding edge theory is the $U(1)/\mathbb{Z}_2$ orbifold, which contains Majorana fermions in its operator content.

\end{abstract}
	
\maketitle

\tableofcontents
	
	
\section{Introduction}

The study of topological phases of matter constitutes one of the most effervescent areas of physics in recent years. Among the reasons for this great interest, we highlight their impressive physical properties, their potential in technological applications, and the complexity of their theoretical descriptions, mingling aspects of quantum mechanics, symmetry, and topology into a deep structure, with far-reaching consequences \cite{Chen,Witten,JMoore}.	
			
One remarkable accomplishment of the field of topological phases of matter is the provision of a concrete platform for the realization of systems exhibiting the elusive Majorana fermions, namely, the topological superconductors (TSs). In recent years, many efforts have been devoted to the study of such systems \cite{Kane,Law,Tanaka,Qi,Chung,Mourik,Liu,Perge,Wang,He}. In addition to their intrinsic interest, the Majorana fermions are generally believed to play a crucial role in quantum computing due to their non-Abelian braiding properties \cite{Kitaev,Simon,Sarma} (for a recent review, see \cite{Marra}). 

A simple setting for the realization of a topological superconductor was proposed in \cite{Qi}, where such phase arises from a quantum anomalous Hall (QAH) phase in proximity to a pairing potential inducing the formation of Cooper pairs (similar models were discussed in \cite{Kane,Wang} and possible experimental signatures were reported in \cite{He}). The specific model of \cite{Qi}, which we shall review in the next section, consists of spinful noninteracting electrons with a {\it nonrelativistic} dispersion relation $E^2= b_1^2 \vec{p}^2 + (\Delta + m_0+ b_2 \vec{p}^2)^2$, where $b_1$ and $b_2$ are positive parameters, $m_0$ is the insulator gap that can be positive or negative, and $\Delta$ is the strength of the pairing potential which is chosen to be positive. The relation between $m_0$ and $\Delta$ dictates in which phase the system is. For weak coupling, $\Delta< |m_0|$, the system is either in a quantum Hall ($m_0<0$) or in a trivial phase ($m_0>0$), but for strong enough coupling, $\Delta > |m_0|$, an $s$-wave topological superconducting phase emerges between them, hosting Majorana fermions both in the vortices and at the edges.

In this paper we revisit this system from the perspective of {\it effective field theory}. Our main objective is to derive a low-energy action incorporating all the phases discussed above. Some interesting aspects concerning the effective field theory are the following.  First, in the pure quantum Hall phase, $\Delta=0$, the effective field theory is expected to be a simple Chern-Simons (CS) action with integer level. Starting with fermions possessing a global $U(1)$ symmetry associated with charge conservation, we follow the usual procedure of introducing a background gauge field for this global symmetry and then integrate out the fermions. This produces a fermionic determinant, which can be computed in the large mass (gap) limit. In relativistic fermionic theories (which correspond to $b_2=0$), the leading contribution in the effective action for the background field is a Chern-Simons term with a half-integer level \cite{Niemi,Redlich}\footnote{In addition to the parity anomaly, this half-integer coefficient would lead to a gauge anomaly, but it can be removed by a local counterterm.}. As our discussion sets itself apart by being nonrelativistic, it is not clear whether a Chern-Simons term can arise from the corresponding fermionic determinant. We carry this computation in this work and show that in fact a Chern-Simons term with a properly quantized level arises from a highly nontrivial combination of several Feynman diagrams, so that we do not need any counterterm to cope with gauge anomaly.  

Another interesting aspect of the effective field theory is evident when we consider the proximity effect to a superconductor. The introduction of a pairing potential in the system breaks the $U(1)$ charge conservation symmetry down to a $\mathbb{Z}_2$ symmetry, allowing the formation of Cooper pairs, so that charge is conserved only mod 2. In this case, we cannot introduce naively continuum gauge fields, since there is no longer a continuous symmetry. We proceed within the Bogoliubov--de Gennes (BdG) formalism (see, for example, \cite{Ludwig,Ryu}), where the fermion operators are accommodated into the Nambu spinors, which are spinors whose components are not independent. This leads to an artificial particle-hole symmetry, such that there is a duplication of the degrees of freedom of the theory. If in addition we work with unconstrained (independent) components in the Nambu spinor, we end up with an enlarged theory possessing fictitious $U(1)$ symmetries \cite{Jackiw}. The physical Hilbert space is then recovered by retaining in the spectrum only states that are properly $\mathbb{Z}_2$ invariants. 

We can take advantage of the $U(1)$ fictitious symmetries in that they can be coupled to background fields, such that we can compute the low-energy effective action again simply by integrating out the fermions. The effective theory is given in terms of Chern-Simons theories, which are locally identical to the one derived in the case of the pure QAH phase. However, the projection onto the physical space amounts to considering the background fields as $O(2)=U(1)\rtimes \mathbb{Z}_2$ gauge fields, instead of $U(1)$. In the $O(2)$ path integral, the sum over nontrivial bundles of $O(2)$ projects out all the states that are not $\mathbb{Z}_2$ invariant. The corresponding edge theory is thus the $U(1)/\mathbb{Z}_2$ orbifold theory, which contains the Majorana fermions.

This work is organized as follows. In Sec. \ref{models}, we review the microscopic models that describe the QAH phase and the TS phase obtained by introducing a pairing potential in the QAH system.  In Sec. \ref{Edge}, we study the edge states of both phases from the point of view of the quantum wires description, where one of the spatial dimensions of the system is discretized. Section \ref{EFT} is dedicated to the derivation of the low-energy effective actions for the QAH and TS phases. In Sec. \ref{orbifold} we study the orbifold edge theory of the TS phase that follows from the effective field theory via bulk-edge correspondence.  We conclude in Sec. \ref{conclusions} with a brief summary. Complementary discussions are presented in Appendices \ref{CH} and \ref{AB}.

	
\section{The Models}\label{models}
	
In this section we briefly review the model introduced in \cite{Qi} describing the transition from the QAH to a TS. When a QAH state is coupled to a conventional $s$-wave superconductor through the proximity effect, the transition between the phases with trivial and nontrivial Hall conductance is, in general, split into two transitions, 
among which appears a chiral TS phase.

	
\subsection{Quantum Anomalous Hall Hamiltonian}
	
The QAH system can be conceived in terms of spinful electrons with a quadratic Hamiltonian, 
\begin{eqnarray}\label{HQAH}
H_{QAH}\equiv\sum_{\vec{p}}\psi^{\dagger}_{\vec{p}}h_{QAH}(\vec{p})\psi_{\vec{p}},
\label{0QAH}
\end{eqnarray}
possessing a $U(1)$ global symmetry corresponding to the charge conservation. The spinor $\psi_{\vec{p}}$ is defined as $\psi_{\vec{p}}\equiv(\begin{array}{cc} c_{\vec{p}\uparrow} & c_{\vec{p}\downarrow}\end{array})^{T}$.  The specific form of the single-particle Hamiltonian $h_{QAH}$ giving rise to the QAH phase is \cite{Qi}
\begin{eqnarray}
h_{QAH}({\vec{p}})\equiv \vec{b}(\vec{p})\cdot \vec{\sigma} = \left(\begin{array}{cc}
m(\vec{p}^2) & b_{1}(p_{x}-ip_{y})\\
b_{1}(p_{x}+ip_{y}) & -m(\vec{p}^2)
\end{array}\right),
\label{QAH}
\end{eqnarray}
with $m(\vec{p}^2) \equiv m_{0} + b_{2}\vec{p}\,^{2}$. The parameters $b_{1}$ and $b_{2}$ are taken to be positive, whereas $m_{0}$ is allowed to change sign, with each one corresponding to a distinct phase. 

To see this, we note that the vector $\vec{b}(\vec{p})=(b_1 p_x, b_1 p_y,m(\vec{p}^2))$ allows us to define the unit vector
\begin{equation}
\vec{n}\equiv \frac{\vec{b}({\vec{p})}}{|\vec{b}(\vec{p})|},
\label{map}
\end{equation}
which in turn corresponds to a map from the momentum region to the unit sphere $S^2$ parametrized by $\vec{n}$. Note that for $|\vec{p}|\rightarrow\infty$, the map \eqref{map} implies $\vec{n}=(0,0,1)$. Therefore, we can think of the momenta $\vec{p}$ as taking values over $\mathbb{R}^2\cup\{\infty\}$, which is topologically equivalent to a sphere $S^2$. In this way, the relation (\ref{map}) defines a class of maps from $S^2$ to $S^2$ classified by the homotopy group $\Pi_2(S^2)=\mathbb{Z}$, with winding number
\begin{equation}
\mathcal{W}(S)=\frac{1}{8\pi} \int_{S^2} d^2p \epsilon^{ij}\vec{n}\cdot (\partial_{p_i}\vec{n}\times\partial_{p_j}\vec{n}).
\end{equation} 
Using the explicit form of the map (\ref{map}), the winding number is
\begin{eqnarray}
\mathcal{W}(S)=\left\{\begin{array}{c}
1\qquad m_{0}<0\\ 0\qquad m_{0}>0.
\end{array}\right.
\label{winding}
\end{eqnarray}

As different values of $\mathcal{W}(S)$ correspond to topologically distinct situations, the system undergoes a phase transition as a function of $m_0$. In this way, the point $m_{0}=0$ is a quantum critical point between a topologically ordered phase when $m_0<0$ (QAH) and a trivial one with $m_0>0$ (trivial insulator). The Hall conductivity $\sigma_{xy}$ is given in terms of the winding number as $\sigma_{xy}=\frac{1}{2\pi}\mathcal{W}$.
	
A hallmark of a topological phase is the presence of gapless edge states when a physical boundary is introduced in the system. This can be incorporated in the Hamiltonian description by promoting $m_0\rightarrow m_0(y)$, with $m_0(y)<0$ if $y<0$ and $m_0(y)>0$ if $y>0$, so that $y=0$ corresponds to an interface between two distinct phases. Hence, there must be edge states at this region. Because of the absence of time-reversal invariance, the edge states are chiral (one-way propagating) and the number of them is related to the bulk topological number (\ref{winding}) through \cite{Witten,Cayssol}
\begin{equation}
\mathcal{W}= N_R-N_L,
\end{equation} 	
where $N_{R/L}$ are the number of chiral right/left propagating edge modes. According to (\ref{winding}), we see that there is a single stable chiral edge mode in the QAH phase.	
	
An elegant way to capture the physics of the edge states is through the bulk-edge correspondence using Chern-Simons effective field theory \cite{Wen}. However, the Lagrangian following from the Hamiltonian \eqref{QAH} is \textit{nonrelativistic}, and the usual procedure of integrating out massive fermions in the presence of a background field is not guaranteed to generate a Chern-Simons term in the effective action. One of the purposes of this work is to show that a Chern-Simons term does emerge in a highly nontrivial way in this nonrelativistic setting. We shall carry out this computation in Sec. \ref{EFT}.

	
\subsection{Proximity Effect to a Topological Superconductor}
	
Next we consider the system in proximity to an $s$-wave superconductor. In this case, a finite pairing amplitude can be induced through the potential $\Delta c^{\dagger}_{\vec{p}\uparrow} c^{\dagger}_{-\vec{p}\downarrow}+\Delta^{\ast} c_{-\vec{p}\downarrow}c_{\vec{p}\uparrow}$, which breaks the $U(1)$ symmetry down to $\mathbb{Z}_2$. The QAH system in the presence of the pairing potential is more conveniently described in terms of the Bogoliubov--de Gennes (BdG) Hamiltonian
\begin{eqnarray}
H_{BdG}=\sum_{\vec{p}}\Psi^{\dagger}_{\vec{p}}h_{BdG}(\vec{p})\Psi_{\vec{p}},
\end{eqnarray}
with the doubled Nambu spinor $\Psi_{\vec{p}}=\left(\begin{array}{cccc}
c_{\vec{p}\uparrow} & c_{\vec{p}\downarrow} & c^{\dagger}_{-\vec{p}\uparrow} & c^{\dagger}_{-\vec{p}\downarrow}
\end{array}\right)^{T}$ and
\begin{eqnarray}
h_{BdG}(\vec{p})=\frac{1}{2}\left(\begin{array}{cc}
h_{QAH}(\vec{p})-\mu & i\Delta\sigma_{y}\\
-i\Delta^{*}\sigma_{y} & -h^{*}_{QAH}(-\vec{p})+\mu
\end{array}\right),
\end{eqnarray}
where $\mu$ is a chemical potential and $\Delta$ is the superconducting gap. If we set $\Delta=0$, this Hamiltonian reduces to (\ref{0QAH}) (with $\mu=0$). In this basis, the spinor $\Psi_{\vec{p}}$ satisfies the constraint
\begin{equation}
\Psi_{\vec{p}}=(\sigma_{x}\otimes\mathbb{1})\Psi^{*}_{-\vec{p}}.
\label{constraint}
\end{equation}
To proceed, we simply ignore this constraint in the intermediate steps, but shall impose it in the end in order to retain the physical space.

In the absence of the chemical potential $\mu$, we can easily block-diagonalize the Hamiltonian, namely,
\begin{eqnarray}
H_{BdG}= \frac{1}{2}\sum_{\vec{p}}\tilde{\Psi}^{\dagger}_{\vec{p}}\left(\begin{array}{cc}
h_{+}(\vec{p}) & 0\\
0 & h_{-}(\vec{p})
\end{array}\right)\tilde{\Psi}_{\vec{p}},
\label{BdG}
\end{eqnarray}
where 
\begin{eqnarray}
h_{\pm}(\vec{p})=\left(\begin{array}{cc}
m(\vec{p}^2)\pm\Delta &  b_{1}(p_{x}-ip_{y})\\
b_{1}(p_{x}+ip_{y}) & -(m(\vec{p}^2)\pm\Delta)
\label{bh}
\end{array}\right),
\end{eqnarray}
and $\tilde{\Psi}_{\vec{p}}=\left(\begin{array}{cccc}
a_{+,\vec{p}} &
a^{\dagger}_{+,-\vec{p}} &
a_{-,\vec{p}} &
- a^{\dagger}_{-,-\vec{p}}			
\end{array}\right)^{T}$, 
with
\begin{equation}
a_{\pm,\vec{p}}\equiv\frac{1}{\sqrt{2}}\left(c_{\vec{p}\uparrow}\pm c^{\dagger}_{-\vec{p}\downarrow}\right). 
\label{newbasis}
\end{equation}
In this basis, the constraint (\ref{constraint}) acts on each block individually,
\begin{equation}
\tilde{\Psi}_{\vec{p}}=(\sigma_z\otimes\sigma_{x})\tilde{\Psi}^{*}_{-\vec{p}}.
\label{constraint1}
\end{equation}		
	
The operators $a_{\pm,\vec{p}}$ do not have a well-defined transformation property under $U(1)$ symmetry. They transform properly only under the subgroup of $U(1)$ transformations $e^{i\alpha}$ for $\alpha=0,\pi$, i.e., under the $\mathbb{Z}_2$ group. Consequently, the excitations created upon application of these operators do not have well-defined electric charge. These considerations are important to correctly identify the physical excitations and, in particular, are extremely useful when we are describing this system in terms of the effective field theory, as it will be done in Sec. \ref{prox}.

\begin{figure}[!h]
\centering
\includegraphics[scale=0.8]{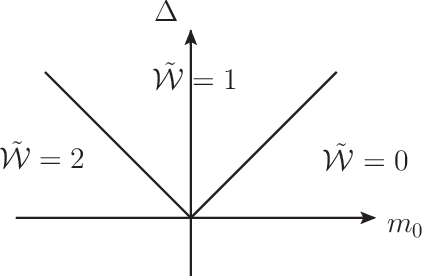}
\caption{Phase diagram of QAH in proximity to a superconducting pairing potential $\Delta$.}
\label{pd}
\end{figure}

Note that the block Hamiltonians in \eqref{bh} are equivalent to two copies of the Hamiltonian \eqref{QAH}, but with a slight modification of the parameters, namely,  $m_0\rightarrow m_0+\Delta$ for the block $h_{+}$ and $m_0\rightarrow m_0-\Delta$ for the block $h_{-}$. Thus, the winding number for the proximity effect can be obtained in the same way as for the QAH system simply as $\widetilde{\mathcal{W}}=\widetilde{\mathcal{W}}_+ + \widetilde{\mathcal{W}}_-$, resulting in
\begin{eqnarray}
\widetilde{\mathcal{W}}=\widetilde{\mathcal{W}}_+ + \widetilde{\mathcal{W}}_-=\left\{\begin{array}{cc}
2& ~~ |m_{0}|>\Delta, ~ \text{with}~ m_0<0\\
1& \!\!\!\!\!\!\!\!\!\!\!\!\!\!\!\!\!\!\!\!\!\!\!\!\!\!\!\!\!\!\!\!\!\!\! |m_{0}|< \Delta\\
0&~~ |m_{0}| >\Delta, ~ \text{with}~ m_0>0.
\end{array}\right.
\label{phases}
\end{eqnarray}
The corresponding phase diagram is shown in Fig. \ref{pd}. 
	
Some comments are in order. In the weak-coupling regime, $|m_0|>\Delta$, with $m_0>0$, the phase is trivial. For $m_0<0$, the system is a nontrivial topological phase which is adiabatically connected to the QAH phase in the limit $\Delta\rightarrow 0$. However, it is important to notice that the winding number here corresponds to the number of gapless chiral fermions in the Majorana basis, because of the particle-hole symmetry inherent to the BdG formalism. So we interpret the two chiral gapless Majorana fermions as glued together to form a complex fermion, which corresponds to the chiral complex mode of the QAH. 
When the pairing becomes stronger, in the region $|m_{0}|< \Delta$, one of the Majorana fermions merges with the bulk states, and we are left with a single Majorana fermion in the boundary. This is the topological superconducting phase. We will see later how these patterns of phase transitions come up in terms of the effective field theory. 
	

\section{Quantum Wires Formulation and Edge States}\label{Edge}

In this section we discuss a simple way to derive the edge states associated with the QAH and TS phases. The idea is to transform the system into a system of quantum wires by discretizing one of the spatial directions. The edge states emerge in a quite natural way in this approach.

	
\subsection{Quantum Anomalous Hall}

The first step is to write a Lagrangian form for the QAH system. The action can be obtained from the Hamiltonian (\ref{0QAH}),
\begin{eqnarray}
S_{QAH}[\psi,\bar{\psi}]=\int d^{3}x\,\bar{\psi}(i\gamma^{0}\partial_0+ib_{1}\gamma^{i}\partial_{i}+b_{2}(i\gamma^{i}\partial_{i})^{2}-m_{0})\psi,
\label{aqah2}
\end{eqnarray}
with the following representation for the Dirac matrices,
\begin{eqnarray}
\gamma^{0}=\sigma^{3}~,~~\gamma^{1}=i\sigma^{2}~~\text{and}~~\gamma^{2}=-i\sigma^{1},
\label{Dirac}
\end{eqnarray}
where $\sigma^{1}, \sigma^2$, and $\sigma^3$ are the Pauli matrices. The spinor $\psi^T=(\psi_{\uparrow}~\psi_{\downarrow})$ is the coordinate counterpart of the fermion operator $\psi_{\vec{p}}$, and the spinor $\bar{\psi}$ is defined in the usual way as $\bar{\psi}\equiv \psi^{\dagger}\gamma^0$. 

With the goal of discretizing the $y$ direction, it is convenient to rotate to a new representation through the unitary operator
\begin{equation}
U=e^{\frac{i \pi}{4} \sigma^2}.
\label{uo}
\end{equation} 
This converts the Dirac matrices to the set
\begin{eqnarray}
\gamma^{0}=-\sigma^{1}~,~~\gamma^{1}=i\sigma^{2}~~\text{and}~~\gamma^{2}=-i\sigma^{3},
\label{DiracNew}
\end{eqnarray}
with the new spinor components
\begin{equation}
\psi_R\equiv \frac{1}{\sqrt{2}}(\psi_{\uparrow}+\psi_{\downarrow})~~~\text{and}~~~\psi_L\equiv \frac{1}{\sqrt{2}}(-\psi_{\uparrow}+\psi_{\downarrow}).
\end{equation}
In this basis, the action in (\ref{aqah2}) becomes
\begin{eqnarray}
	S&=&\int d^3x \left[ i\psi_{R}^{\dagger}\partial_{+}\psi_{R} +i\psi_{L}^{\dagger}\partial_{-}\psi_{L}-b_{2}\psi_{L}^{\dagger}\partial_{x}^{2}\psi_{R}-b_{2}\psi_{R}^{\dagger}\partial_{x}^{2}\psi_{L}\right.\nonumber\\
	&+& \left. m_0 \psi_{L}^{\dagger}\left(1-m_0^{-1}b_{1}\partial_{y}-m_0^{-1} b_{2}\partial_{y}^{2}\right)\psi_{R}
	+m_0 \psi_{R}^{\dagger}\left(1+m_0^{-1}b_{1}\partial_{y}-m_0^{-1} b_{2}\partial_{y}^{2}\right)\psi_{L}\right],
	\label{a01}
\end{eqnarray}
where we have defined $\partial_{\pm}\equiv \partial_0 \pm b_1 \partial_1$. We are assuming that the mass $m_0$ is nonvanishing, so that the bulk remains always gapped. We   discuss first the QAH phase, where the mass $m_0$ is negative. In this case, let us choose a representative point in the phase diagram corresponding to the QAH phase, where the edge states appear in a quite direct way. This is achieved by choosing the parameters so that
\begin{align}
	b_{2}=-\frac{b_{1}^{2}}{2m_0}. 
	\label{finet}
\end{align}  
Notice that, as $b_1$ and $b_2$ are positive, this condition can only be satisfied for negative $m_0$, i.e., in the QAH phase. With this choice, the action can be written as
\begin{eqnarray}
	S&=&\int d^3x\left[ i\psi_{R}^{\dagger}\partial_{+}\psi_{R} +i\psi_{L}^{\dagger}\partial_{-}\psi_{L}-\frac{b_{1}^{2}}{2|m_0|}\psi_{L}^{\dagger}\partial_{x}^{2}\psi_{R}-\frac{b_{1}^{2}}{2|m_0|}\psi_{R}^{\dagger}\partial_{x}^{2}\psi_{L}\right.\nonumber\\
	&-&\left.|m_0| \psi_{L}^{\dagger}\left(1+a\partial_{y}+\frac12 a^2 \partial_{y}^{2}\right)\psi_{R}
	-|m_0 |\psi_{R}^{\dagger}\left(1-a\partial_{y}+\frac12 a^2\partial_{y}^{2}\right)\psi_{L}\right],
	\label{a03}
\end{eqnarray}
where we have introduced the ``wire spacing" $a\equiv \frac{b_1}{|m_0|}$. For a large gap $|m_0|$, the wire spacing $a$ is small and the terms in the second line of this expression can be identified as second-order Taylor expansions, so that (\ref{a03}) is approximated by 
\begin{eqnarray}
	S&\approx& \int d^3x \left[ i\psi_{R}^{\dagger}\partial_{+}\psi_{R} +i\psi_{L}^{\dagger}\partial_{-}\psi_{L}-\frac{a b_{1}}{2}\psi_{L}^{\dagger}\partial_{x}^{2}\psi_{R}-\frac{a b_{1}}{2}\psi_{R}^{\dagger}\partial_{x}^{2}\psi_{L}\right.\nonumber\\
	&-&\left.\frac{b_1}{a}\psi_{L}^{\dagger}(t,x,y)\psi_{R}(t,x,y+a) -\frac{b_1}{a}\psi_{R}^{\dagger}(t,x,y+a)\psi_{L}(t,x,y)\right].
	\label{a02}
\end{eqnarray}
This form leads naturally to the discretization of the $y$ direction. To this, we make the following prescriptions:
\begin{equation}
\psi_{R/L}(t,x,y)\rightarrow\frac{1}{\sqrt{a}}\psi^{j}_{R/L}(t,x) \qquad\text{and} \qquad \int dy \rightarrow a \sum\limits_{j}. 
\label{discretization}
\end{equation}
Then, by considering the system with open boundary conditions in the $y$ direction and using (\ref{discretization}), the action (\ref{a02}) becomes
\begin{eqnarray}
	S& \approx& \int d^2x \left[ \sum_{j=1}^{N} \left(i\psi_{R}^{j\dagger}\partial_{+}\psi_{R}^{j} +i\psi_{L}^{i\dagger}\partial_{-}\psi_{L}^{j}
	\right) - \frac{b_1}{a}\sum \limits_{j=1}^{N-1}\left(\psi_{L}^{j\dagger}\psi_{R}^{j+1}+\psi_{R}^{j+1 \dagger}\psi_{L}^{j}\right)+\cdots\right],
\end{eqnarray}
where we have discarded the irrelevant terms of order $a$. We notice that the chiral modes $\psi_R^1$ and $\psi_L^N$, associated with the first and the last wires, respectively, are decoupled from the remaining modes in this action. In fact, they are governed exclusively by kinetic terms 
\begin{equation}
S_{\text{edge}}= \int d^2x \left(i\psi_{R}^{1\dagger}\partial_{+}\psi_{R}^{1} + i\psi_{L}^{N\dagger}\partial_{-}\psi_{L}^{N} \right),
\end{equation}
and thus are identified as the gapless edge states of the QAH phase. It is worth emphasizing that these modes are spatially far apart from each other and, due to the locality of the interactions, they cannot be gapped through backscattering. Furthermore, deviations from (\ref{finet}) will generically introduce interactions between the gapless edge modes and gapped modes both at the same wire and at the neighboring wires. However, as long as the interaction strength is small compared to the gap, the edge states will remain gapless. These features imply that they are stable. In sum, we have found a complex one-way propagating fermion in each one of the boundaries; i.e., the edge theory of the QAH phase is a conformal field theory with chiral central charge $c=1$.

Now we discuss the system in the trivial phase $m_0>0$. In this case, we cannot proceed as in the previous one, since the terms in the second line of \eqref{a01}, namely,
\begin{equation}
\psi_{L}^{\dagger}\left(1-m_0^{-1}b_{1}\partial_{y}-m_0^{-1} b_{2}\partial_{y}^{2}\right)\psi_{R}+ \text{H. c.},
\end{equation}
do not  correspond to a second-order Taylor expansion for any choice of the parameters and, consequently, after discretization no mode will be left decoupled. To see this, we define again the ``wire spacing" $a\equiv \frac{b_1}{m_0} $ so that, for large gap, the above expression can be written as
\begin{eqnarray}
&&\psi_L^{\dagger}(t,x,y)\psi_R(t,x,y-a) -\frac{b_2}{a b_1}\psi_L^{\dagger}(t,x,y) \left[\psi_R(t,x,y+a)\right.\nonumber\\
&-& \left. 2\psi_R(t,x,y)+\psi_R(t,x,y-a) \right]+ \text{H. c.}.
\end{eqnarray}
We then proceed with the discretization of these terms according to the prescriptions in (\ref{discretization}), taking into account open boundary conditions in the $y$ direction. This produces in the action terms proportional to 
\begin{eqnarray}
&-&\frac{ b_2}{a b_1}\sum_{j=1}^{N-1} \psi_L^{j \dagger}(t,x)\psi_R^{j+1}(t,x)+ \frac{2 b_2}{a b_1}\sum_{j=1}^{N} \psi_L^{j \dagger}(t,x)\psi_R^{j}(t,x)\nonumber\\
&+&\left(1- \frac{b_2}{a b_1}\right)\sum_{j=2}^{N} \psi_L^{j \dagger}(t,x)\psi_R^{j-1}(t,x)+\text{H. c.},
\end{eqnarray}
which shows that all modes partake in the interaction (for any choice of the parameters) and then turn out to be gapped. There are no gapless edge states left behind, as expected for the topologically trivial phase.


\subsection{Superconducting Phase}

Now we consider the system in the presence of the superconducting pairing potential and move to the BdG formalism. The action corresponding to the Hamiltonian (\ref{BdG}) is
\begin{eqnarray}
	S_{QAH-TS}&\equiv& \frac12 \int d^3x \left[\mathcal{L}_{+}(\psi_{+},\bar{\psi}_{+}) + \mathcal{L}_{-}(\psi_{-},\bar{\psi}_{-})\right]\nonumber\\
	&=&	\frac{1}{2}\int d^{3}x \left[ \bar{\psi}_+(i\gamma^{0}\partial_0+ib_{1}\gamma^{i}\partial_{i}+b_{2}(i\gamma^{i}\partial_{i})^{2}-(m_{0}+\Delta))\psi_+\right.\nonumber\\
	&+&\left. \bar{\psi}_-(i\gamma^{0}\partial_0+ib_{1}\gamma^{i}\partial_{i}+b_{2}(i\gamma^{i}\partial_{i})^{2}-(m_{0}-\Delta))\psi_-\right],
	\label{supphase}
\end{eqnarray}
where the doubled spinors are $ \psi_{\pm}^{T}=\left(\begin{array}{cc}
	a_{\pm}&a_{\pm}^{\dagger} 
\end{array} \right)$. 
We can then follow a strategy similar to the previous case for discretizing this action.

The first step is to rotate according to the unitary operator in (\ref{uo}), under which the spinors transform as 
\begin{align}
	\psi_{\pm} ~\rightarrow ~\tilde{\psi}_{\pm} \equiv U \psi_\pm=\frac{1}{\sqrt{2}}\left(\begin{array}{cc}
	a_{\pm}+a_{\pm}^{\dagger}	& a_{\pm}^{\dagger}-a_{\pm}
	\end{array}\right)^{T}.
\end{align}
The transformed spinors naturally lead to the introduction of the Majorana operators
\begin{equation}
	\chi_{R}^{\pm}\equiv\frac{1}{\sqrt{2}}\left(a_{\pm}+a_{\pm}^{\dagger}\right) ~~~\text{and}~~~ \chi_{L}^{\pm}\equiv \frac{i}{\sqrt 2}\left(a_{\pm}-a_{\pm}^{\dagger}\right),
\end{equation}
so that 
\begin{equation}
\tilde{\psi}_{\pm}=(\chi_{R}^{\pm} ~ i \chi_{L}^{\pm})^T.
\end{equation}
In terms of the Majorana fermions, the corresponding Lagrangians become
\begin{eqnarray}
	\mathcal{L}_{\pm}&=&i \chi_{R}^{\pm}\partial_{+}\chi_{R}^{\pm}+i \chi_{L}^{\pm}\partial_{-}\chi_{L}^{ \pm}+2 i b_{2}\chi_{L}^{\pm}\partial_{x}^{2}\chi_{R}^{\pm}
	\nonumber\\
	&-&2 i m_{\pm}\chi_{L}^{\pm}\left(1-m_{\pm}^{-1}b_{1}\partial_{y}-m_{\pm}^{-1}b_{2}\partial_{y}^{2}\right)\chi_{R}^{\pm},
\end{eqnarray}
where we have defined the masses $ m_{\pm}\equiv m_{0}\pm \Delta$. 

We are mostly interested in studying the edge states of the superconducting phase, where $m_+>0$ and $m_-<0$. In this situation, the contribution of $\mathcal{L}_+$ works precisely as the Lagrangian of the trivial phase discussed previously, and consequently does not lead to any massless edge state. On the other hand, the contribution of $\mathcal{L}_-$ works like the Lagrangian of the QAH phase, but with the difference that now the edge states are given in terms of Majorana fermions. In fact, we see that the term
\begin{equation}
-2 i m_{-}\chi_{L}^{-}\left(1-m_{-}^{-1}b_{1}\partial_{y}-m_{-}^{-1}b_{2}\partial_{y}^{2}\right)\chi_{R}^{-}
\label{p}
\end{equation}
of $\mathcal{L}_-$ can be identified as a second-order Taylor expansion provided we choose the representative point
\begin{equation}
b_2= -\frac{b_1^2}{2 m_-}.
\end{equation}
With this choice and identifying the wire spacing as $a\equiv \frac{b_1}{|m_{-}|}$, the expression (\ref{p}) can be approximated by
\begin{equation}
\frac{2 i b_1}{a} \chi_L^{-}(t,x,y) \chi_R^{-}(t,x,y+a),
\end{equation}
in the limit of large gap $|m_-|$. After the discretization using the prescriptions in (\ref{discretization}), this leads to a term in the action proportional to
\begin{equation}
\sum_{j=1}^{N-1}  \chi_{L,j}^{-}(t,x)\chi_{R,j+1}^{- }(t,x),
\end{equation}
which implies that the Majorana modes $\chi_{R,1}^-$ and $\chi_{L,N}^-$, associated with the first and the last wires, respectively, are decoupled and then remain gapless. Therefore, the edge theory of the superconducting phase is given by
\begin{equation}
S_{\text{edge}}= \int d^2x \left(i\chi_{R,1}^{-}\partial_{+}\chi_{R,1}^{-} + i\chi_{L,N}^{-}\partial_{-}\chi_{L,N}^{-} \right),
\end{equation}
which corresponds to a conformal field theory with chiral central charge $c=\frac12$. 

We shall return to the edge theory later on, when discussing how the edge states can be recovered from the effective field theory through the bulk-edge correspondence.


\section{Effective Field Theory}\label{EFT}
		
The main goal of this section is to derive the low-energy effective field theory for the QAH system in proximity to an $s$-wave superconductor. Given the nonrelativistic character of the fermion Lagrangian, it is not clear whether a Chern-Simons term can arise from the corresponding fermionic determinant.

		
\subsection{Topological Effective Field Theory for the QAH Phase}
		
The nontrivial part of the computation of the effective action is already present in the case of the pure QAH system, and so we will concentrate first on this case. The strategy is to introduce a background gauge field $A$ for the global $U(1)$ symmetry of the QAH, and then integrate out the gapped fermions to obtain $S_{eff}[A]$.

The corresponding action is given in (\ref{aqah2}), which we repeat here for convenience,
\begin{eqnarray}
S_{QAH}[\psi,\bar{\psi}]=\int d^{3}x\,\bar{\psi}(i\gamma^{0}\partial_0+ib_{1}\gamma^{i}\partial_{i}+b_{2}(i\gamma^{i}\partial_{i})^{2}-m_{0})\psi.
\label{aqah}
\end{eqnarray}
Then we introduce a background gauge field $A$ for the global $U(1)$ symmetry\footnote{Some aspects of this model with dynamical gauge field and positive mass have been analyzed previously in \cite{Charneski}.},
\begin{eqnarray}
S_{QAH}[\psi,\bar{\psi};A]=\int d^{3}x\,\bar{\psi}(i\gamma^{0}D_{0}+ib_{1}\gamma^{i}D_{i}+b_{2}(i\gamma^{i}D_{i})^{2}-m_{0})\psi,
\label{aqahA}
\end{eqnarray}
where $D_{\mu}\equiv\partial_{\mu}-iA_{\mu}$, $\mu=0,1,2$. Before proceeding, it seems that there is certain ambiguity in this process. Indeed, if we write the term with coefficient $b_2$ in (\ref{aqah}) as $b_{2}(i\gamma^{i}\partial_{i})^{2}=b_2 \vec{\nabla}^2$, then the gauging of this term is simply $b_2 \vec{D}^2$, which is different from the term $b_{2}(i\gamma^{i}D_{i})^{2}$ of (\ref{aqahA}), namely,
\begin{eqnarray}
(i\gamma^{i}D_{i})^{2}&=& \vec{D}^2 + \frac{i }{4}[\gamma^i,\gamma^j]F_{ij}\nonumber\\
&=& \vec{D}^2 - \frac{ 1}{2}\gamma^0 \epsilon^{0ij} F_{ij},
\label{ambiguity}
\end{eqnarray}
where $F_{ij}=\partial_i A_j - \partial_j A_i$ and we have used $[\gamma^{\mu},\gamma^{\nu}]=-2i\epsilon^{\mu\nu\rho}\gamma_{\rho}$. Therefore, these two ways of treating the higher-derivative term produce theories differing by an operator proportional to $\bar{\psi}\gamma^0 \epsilon^{0ij} F_{ij}\psi$. In general, such a UV operator is expected simply to renormalize the parameters of the low-energy effective theory. However, as we shall discuss later, it does not affect the induced topological Chern-Simons term since it amounts just to a redefinition of the current. In the following we proceed with the form (\ref{aqahA}).

The strategy is to compute the effective action for the background gauge field $A$ by integrating out the fermions,
\begin{equation}
e^{i S_{eff}[A]}= \int\mathcal{D}\psi\mathcal{D}\bar\psi\text{e}^{i S_{QAH}[\psi,\bar{\psi}]}.
\end{equation}
The effective action is then organized in a local expansion in powers of the external field, 
\begin{eqnarray}
i S_{eff}[A]&=&\int d^3x \frac{1}{2}\left(A_0 \Pi_{00} A_0 + A_0 \Pi_{0i}A_i+A_i \Pi_{i0}A_0 + A_i \Pi_{ij} A_j +\cdots  \right)\nonumber\\
&=&\int d^3x \frac{1}{2}\left(A_0 \Pi_{00} A_0 + 2 A_0 \Pi_{0i}A_i + A_i \Pi_{ij} A_j +\cdots  \right),
\label{effec}
\end{eqnarray}
where we have used the fact that the two terms involving $A_0$ and $A_i$ give the same contribution up to integration by parts. The operators $\Pi_{\mu\nu}$ can be computed from the relevant Feynman diagrams with fermions in the internal lines and the background field in the external ones.  
		
\begin{figure}[!h]
\includegraphics[scale=0.6]{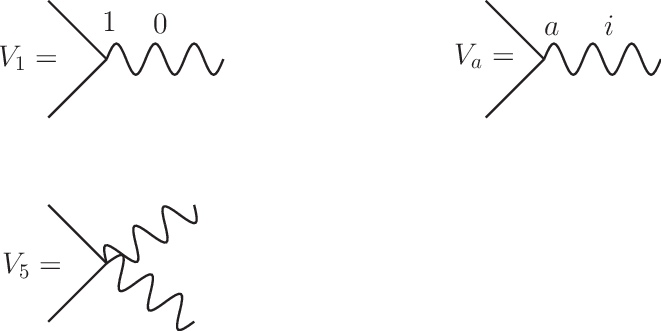}
\caption{Interaction vertices. The index 0 in the diagram $V_1$ means that the photon line involves the component $A_0$. The diagram $V_a$ represents generically the three vertices with $a=2,3,4$, and the index $i$ means that the photon line involves the component $A_i$.  }
\label{fr}
\end{figure}

\begin{figure}[!h]
\includegraphics[scale=0.7]{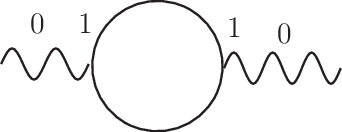}
\caption{One-loop contribution for the two-point function $\langle A_0 A_0 \rangle$, which contributes for $\Pi_{00}$ in the effective action (\ref{effec}). }
\label{A0A0}
\end{figure}

\begin{figure}[!h]
\includegraphics[scale=0.7]{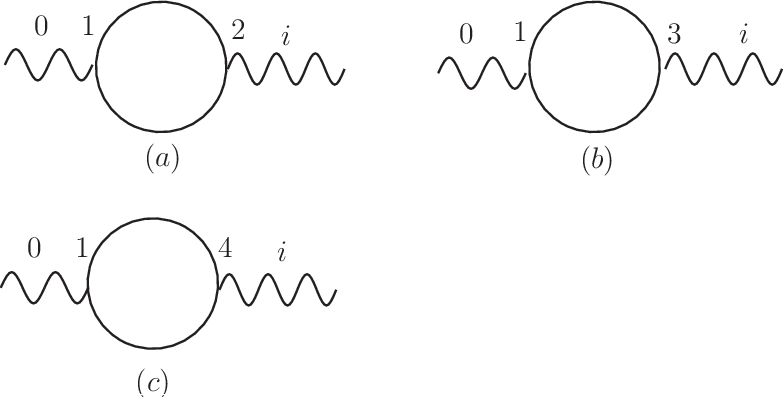}
\caption{One-loop contributions for the two-point function $\langle A_0 A_i \rangle$, which contribute for $\Pi_{0i}$ in the effective action (\ref{effec}). }
\label{A0Ai}
\end{figure}

\begin{figure}[!h]
\includegraphics[scale=0.7]{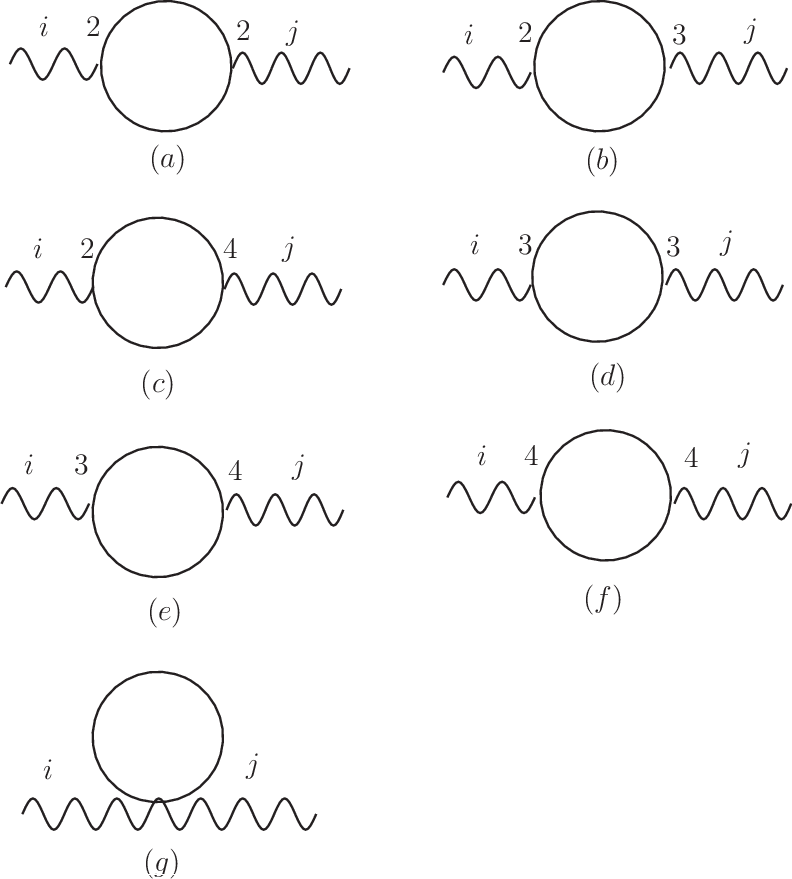}
\caption{One-loop contributions for the two-point function $\langle A_i A_j \rangle$, which contribute for $\Pi_{ij}$ in the effective action (\ref{effec}). }
\label{AiAj}
\end{figure}

From (\ref{aqahA}) we can identify immediately the interaction vertices,
\begin{equation}
V_1 \equiv \bar{\psi}\gamma^0A_0\psi,~~~V_2 \equiv  b_1\bar{\psi}\gamma^iA_i\psi,
\end{equation}
and
\begin{equation}
V_3 \equiv -i b_2 (\partial_{i}\bar{\psi})\gamma^{i}\gamma^{j}A_{j}\psi, ~~~V_4 \equiv ib_{2}\bar{\psi}\gamma^{i}\gamma^{j}A_i\partial_{j}\psi,~~~V_5 \equiv b_{2}\bar{\psi}\left(\gamma^{i}A_{i}\right)^{2}\psi.
\end{equation}
They are shown in Fig. \ref{fr}. The fermion propagator is
\begin{eqnarray}
S(k)&=&\frac{i}{\gamma^{0}k_{0}-b_{1}\vec{\gamma}\cdot\vec{k}-b_{2}\vec{k}^2-m_{0}+i \epsilon}\nonumber\\
&=&  \frac{i(k_0\gamma^0-b_{1}\vec{k}\cdot \vec{\gamma}+b_2 \vec{k}^2+m_0)}{k_0^2-b_1^2 \vec{k}^2-(b_2 \vec{k}^2+m_0)^2+i\epsilon}.
\end{eqnarray}
The one-loop contributions to the two-point functions of the background field are shown in Figs. \ref{A0A0}, \ref{A0Ai}, and \ref{AiAj}. 

The one-loop contributions are organized in a derivative expansion or, equivalently, in a momentum expansion. There are no zero-th-order contributions to the two-point function $\langle A_{\mu}A_{\nu} \rangle$. 

We then consider the first-order derivative contributions to the two-point function $\langle A_0A_i\rangle$. They are contained in the diagrams of Fig. \ref{A0Ai}, and furnish
\begin{equation}
\Pi_{0i}^{(a)}(p)=\left[\Theta(m_0)\left(\frac{-b_1^2}{4\pi(b_1^2+4b_2 m_0)}\right)+\Theta(-m_0)\left(\frac{1}{4\pi} \right)\right] \epsilon_{0ij}p_j+\cdots,
\end{equation}
\begin{equation}
\Pi_{0i}^{(b)}(p)=\Pi_{0i}^{(c)}(p)=\left[\Theta(m_0)\left(\frac{b_1^2}{8\pi(b_1^2+4b_2 m_0)}\right)+\Theta(-m_0)\left(\frac{1}{8\pi} \right)\right] \epsilon_{0ij}p_j+\cdots,
\end{equation}
where $\Theta$ is the Heaviside step function. Adding up the three pieces, we get
\begin{equation}
\Pi_{0i}(p)=\Pi_{0i}^{(a)}+\Pi_{0i}^{(b)}+\Pi_{0i}^{(c)}= \Theta(-m_0) \frac{1}{2\pi} \epsilon_{0ij}p_j+\cdots.
\end{equation}
		
Finally, we consider the two-point function $\langle A_iA_j\rangle$.  Only the diagrams $(a)$, $(b)$, and $(c)$ of Fig. \ref{AiAj} contribute to the first-order derivative term. The corresponding contributions are
\begin{equation}
\Pi_{ij}^{(a)}(p)=\Theta(m_0)\left(\frac{-b_1^2}{2\pi(b_1^2+4b_2 m_0)}\right)\epsilon_{0ij}p_0+\cdots,
\end{equation}
\begin{equation}
\Pi_{ij}^{(b)}(p)=\Pi_{ij}^{(c)}(p)=\left[\Theta(m_0)\left(\frac{b_1^2}{4\pi(b_1^2+4b_2 m_0)}\right)+\Theta(-m_0)\left(\frac{1}{4\pi} \right)\right] \epsilon_{0ij}p_0+\cdots.
\end{equation}
Adding the three terms, we obtain
\begin{equation}
	\Pi_{ij}^{(a)}+\Pi_{ij}^{(b)}+\Pi_{ij}^{(c)}= \Theta(-m_0) \frac{1}{2\pi} \epsilon_{0ij}p_0+\cdots.
\end{equation}
The diagrams $(d)$, $(e)$, and $(f)$ contribute at least with two derivatives, while the diagram $(g)$ does not contribute at all.
		
Including all the first-order contributions to the low-energy effective action (\ref{effec}), we obtain
\begin{eqnarray}
i S_{eff}[A]&=&\int \!\!\frac{d^3p}{(2\pi)^3} \left[ \Theta(-m_0) A_0(p)\left(   \frac{1}{2\pi} \epsilon_{0ij}p_j  \right)  A_i(-p) + \Theta(-m_0)  A_i(p)\left(   \frac{1}{4\pi} \epsilon_{0ij}p_0  \right)  A_j(-p)  +\cdots\right]\nonumber\\
&=&\int \! \! \frac{d^3p}{(2\pi)^3} \left[ - \Theta(-m_0) \frac{1}{4\pi}\epsilon^{\mu\nu\rho}A_{\mu}(p)p_{\nu}A_{\rho}(-p)+\cdots \right].
\end{eqnarray} 
In the coordinate space this reads
\begin{equation}
S_{eff}[A]= \int d^3x \left[\Theta(-m_0) \frac{1}{4\pi} \epsilon^{\mu\nu\rho}A_{\mu}\partial_{\nu}A_{\rho}+\cdots \right].
\label{eacs0}
\end{equation}
It is quite remarkable that even though the fermion dynamics is nonrelativistic, a usual Chern-Simons term is recovered in the topological sector of the effective theory. 
We can further appreciate this result recalling that in any gapped system of charged fermions the Hall conductivity, $\sigma_{xy}$, is quantized, and it is given in terms of the winding number of the momentum space Hamiltonian \cite{Wu12}. The Hall responses $J^i=\sigma_{xy} \epsilon^{ij}E^j$  and $\sigma_{xy}=\frac{\partial J^0}{\partial B}$ (Streda formula) to the application of electric and magnetic fields in the system, can be combined in a covariant way as  $J^{\mu}=\sigma_{xy} \epsilon^{\mu\nu\rho}\partial_{\nu}A_{\rho}$. 
Then, these responses can be obtained from the coupling $A_{\mu}J^{\mu}$, which in terms of an effective action corresponds to $\frac{\sigma_{xy}}{2}\int d^3x \epsilon^{\mu\nu\rho}A_{\mu}\partial_{\nu}A_{\rho}$. This is an indication that, even though the fermion system is nonrelativistic, the field theory has to manage to deliver a usual CS term, whose coefficient is related to the Hall conductivity.

In the present case, the Chern-Simons term arises from a nontrivial combination of several one-loop diagrams, whereas in the usual (relativistic) case the Chern-Simons term comes from a single one-loop diagram. The higher-order derivative corrections represented by the dots are nonrelativistic. In addition, the CS coefficient in (\ref{eacs0}) is properly quantized and does not break invariance under large gauge transformations, in contrast to the usual case of the relativistic fermionic determinant. The Hall conductivity $\sigma_{xy}$ can be read from the Chern-Simons coefficient $\frac{\sigma_{xy}}{2}\int AdA$, which implies that $\sigma_{xy}=\frac{\Theta(-m_0)}{2\pi}$. This recovers precisely the results of the previous section for the Hall conductivity $\sigma_{xy}=\frac{1}{2\pi}\mathcal{W}$, with $\mathcal{W}$ given in (\ref{winding}). In Appendix \ref{CH}, we discuss that the CS term is protected against radiative corrections when the gauge field is dynamical.

Now we discuss that the introduction of a chemical potential $\mu$ in the system does not affect the topological sector of the effective field theory. The introduction of the chemical potential amounts to replacing $h_{QAH}(\vec{p})\rightarrow h_{QAH}(\vec{p}) -\mu$ in the Hamiltonian (\ref{0QAH}). This, in turn, corresponds to the presence of the term $\mu\bar{\psi}\gamma^0\psi$ in the Lagrangian (\ref{aqah}). We can absorb this factor in the time derivative, $\tilde{\partial}_0\equiv \partial_0-i\mu$, or equivalently, in the zero component of the momentum, $\tilde{p}_0\equiv p_0+\mu$. Therefore, it is immediate to see that the chemical potential does not affect the Chern-Simons contribution, 
\begin{equation}
\int d^3p A_{i}(p) \epsilon_{0ij} \tilde{p}_0 A_j(-p)= \int d^3p A_{i}(p) \epsilon_{0ij} p_0 A_j(-p),
\end{equation}
since $\mu \int d^3p A_{i}(p) \epsilon_{0ij}A_j(-p)$ vanishes by symmetry.

Before move on, we discuss the effect of considering $\vec{D}^2$ instead of $(i\gamma^iD_i)^2$ in the definition of the QAH system in the presence of an external field.
According to Eq. (\ref{ambiguity}), this amounts to including in the previous computation a new interaction vertex that we denote as
\begin{equation}
V_6\equiv \frac{b_2}{2}\bar{\psi}\gamma^0\epsilon_{0ij}F_{ij}\psi.
\end{equation}
This new vertex has the potential to generate a Chern-Simons term in combination with the vertices $V_1$, $V_2$, $V_3$, and $V_4$. However, we can check explicitly that they do not generate any Chern-Simons contribution, so that the topological sector of the theory is not sensitive to the presence of the above operator. 

To understand the underlying reason for this, we consider the $U(1)$ current following from \eqref{aqahA}. It can be easily constructed by considering the generalization of the Noether theorem for higher derivative theories (see for example \cite{Gomes}). In particular, for an internal symmetry the current for a second-order derivative theory reads
\begin{equation}
J^{\mu}=\frac{\partial\mathcal{L}}{\partial \partial_{\mu}\phi_I}\delta\phi_I + \frac{\partial\mathcal{L}}{\partial \partial_{\mu}\partial_{\nu}\phi_I}\partial_{\nu}\delta\phi_I -
\partial_{\nu}\left(\frac{\partial\mathcal{L}}{\partial \partial_{\mu}\partial_{\nu}\phi_I}\right) \delta\phi_I,
\end{equation}
where $\phi_I$ stands for a generic set of fields. Using this expression for the global $U(1)$ symmetry of \eqref{aqahA}, we obtain the components
\begin{equation}
J^{0}=\bar{\psi}\gamma^{0}\psi~~~\text{and}~~~J^{i} = b_{1}\bar{\psi}\gamma^{i}\psi+i b_{2}(\bar{\psi}\gamma^{i}\gamma^{j}\partial_{j}\psi-\partial_{j}\bar{\psi}\gamma^{j}\gamma^{i}\psi)+2 b_2 \bar{\psi}\psi A^i.
\label{0c}
\end{equation}
A conserved current is not uniquely defined, since we can always redefine it as
\begin{equation}
J^{\prime\mu}=J^{\mu}+\partial_{\nu}\Omega^{\mu\nu},~~~\text{with}~~~ \Omega^{\mu\nu}=-\Omega^{\nu\mu}.
\label{current}
\end{equation}
The new current $J^{\prime\mu}$ is as good as the initial one, once it is also conserved and gives rise to the same charge $\int_{\text{space}} J^{0}$. 

Next we consider the current in the theory with the operator $b_2 \vec{D}^2$ instead of $b_2(i\gamma^iD_i)^2$. In this case,  the current is given by
\begin{equation}
J^{\prime 0}= J^0~~~\text{and}~~~J^{\prime i}=J^i - b_2 \partial_{j}\left(\bar{\psi}\epsilon^{ij0}\gamma_0\psi\right). 
\end{equation}
We see that the currents $J^{\mu}$ and $J^{\prime \mu}$ differ precisely by a term of the form \eqref{current}, with $\Omega^{0i}=0$ and $\Omega^{ij}=- b_2\bar{\psi}\epsilon^{ij0}\gamma_0\psi$.

Next, we consider the generic relation
\begin{equation}
\frac{\delta S_{eff}[A]}{\delta A_{\mu}}=\langle J^{\mu} \rangle.
\label{ea1}
\end{equation}
Denoting by $S_{eff}^{\prime}[A]$ the effective action coming from the UV theory involving $b_2 \vec{D}^2$ instead of $b_2(i\gamma^iD_i)^2$, we have likewise
\begin{equation}
\frac{\delta S^{\prime}_{eff}[A]}{\delta A_{\mu}}=\langle J^{\prime\mu} \rangle.
\label{ea2}
\end{equation}
Now consider the density in the relations \eqref{ea1} and \eqref{ea2}. As $J^0=J^{\prime 0}$, it follows that
\begin{equation}
\frac{\delta}{\delta A_0}\left(S_{eff}^{\prime}[A]- S_{eff}[A]\right)=0,
\end{equation}
namely, the corresponding effective actions differ only by gauge-invariant terms independent of $A_0$,
\begin{equation}
S_{eff}^{\prime}[A]= S_{eff}[A]+\int d^3x \, \mathit{O}(A_i).
\end{equation}
Therefore, the CS contribution must be exactly the same in both effective actions.



\subsection{EFT for the QAH Phase in Proximity to a TS}\label{prox}

After the derivation of the effective field theory for the QAH system, we are ready to study the effective field theory in the presence of the superconducting pairing potential. As discussed previously, in this case the $U(1)$ global symmetry is broken down to the discrete $\mathbb{Z}_2$ symmetry. However, as is usual in treating superconductor systems, it is quite useful to insist on keeping a {\it fictitious} $U(1)$ symmetry. To conceive this, we simply ignore the constraint (\ref{constraint1}) and work with unconstrained spinors. This means that for each block in (\ref{BdG}) we have a fictitious $U(1)$ symmetry. 

In quantizing the theory with the fictitious $U(1)$ global symmetries we find that the states in this enlarged Hilbert space are in the representation of the fictitious $U(1)$. Among all these states, we have to project out all the states which are not invariant under the action of $\mathbb{Z}_2$, so that the physical Hilbert space contains only states that are properly $\mathbb{Z}_2$ invariant.

We can take advantage of the fictitious $U(1)$ symmetries to introduce background fields for them. In this way, we can follow the strategy of the previous section and integrate out the fermions to derive the (local) effective action for these background fields. The above projection onto the physical Hilbert space can be implemented concretely by considering the background fields as actually $O(2)=U(1)\rtimes \mathbb{Z}_2$ gauge fields, instead of $U(1)$. The local action is precisely the action of a $U(1)\times U(1)$  theory but the corresponding path integrals are different.  In the $O(2)$ case we have to sum over the $\mathbb{Z}_2$ bundles, which is equivalent to computing the path integral with the insertion of projection operators that select only the $\mathbb{Z}_2$-invariant states. 
		
We start considering again the action (\ref{supphase}), 
\begin{eqnarray}
S_{QAH-TS}&=& \frac{1}{2}\int d^{3}x \left[ \bar{\psi}_+(i\gamma^{0}\partial_0+ib_{1}\gamma^{i}\partial_{i}+b_{2}(i\gamma^{i}\partial_{i})^{2}-(m_{0}+\Delta))\psi_+\right.\nonumber\\
&+&\left. \bar{\psi}_-(i\gamma^{0}\partial_0+ib_{1}\gamma^{i}\partial_{i}+b_{2}(i\gamma^{i}\partial_{i})^{2}-(m_{0}-\Delta))\psi_-\right],
\end{eqnarray}
where $\psi_{\pm}$ are the unconstrained two-component spinors associated with the two blocks of (\ref{BdG}). Upon the constraint (\ref{constraint1}), we get in the momentum space simply
\begin{equation}
\psi_{+}^T(\vec{p})=(a_{+,\vec{p}},a_{+,-\vec{p}}^{\dagger})~~~\text{and}~~~\psi_{-}^T(\vec{p})=(a_{-,\vec{p}} ,-a^{\dagger}_{-,-\vec{p}}).
\end{equation}
		
Proceeding with the unconstrained spinors, we can then introduce the corresponding background fields $A^{(\pm)}$ and compute the low-energy effective theory by integrating out the fermions, exactly as we did in the previous section. One subtle point that arises in this procedure is that there is no natural way to assign fictitious $U(1)$ charges for $\psi_{\pm}$. Therefore, we shall keep them unfixed momentarily, which amounts to considering covariant derivatives $D_{\mu}= \partial_{\mu}- i q A_{\mu}^{(\pm)}$, for a generic integer charge $q$. We will fix the charge posteriorly on physical grounds, so that the resulting effective theory describes properly the underlying physics.

The effective action in this case can be immediately read out from the result (\ref{eacs}), with the appropriate adjustment of the parameters, as well as a corresponding rescaling of gauge fields by a factor of $q$,
\begin{eqnarray}
S_{eff}[A^{(+)},A^{(-)}]&=& \int d^3x \left[\Theta(-m_0-\Delta) \frac{q^2}{4\pi} \epsilon^{\mu\nu\rho}A^{(+)}_{\mu}\partial_{\nu}A^{(+)}_{\rho}\right. 
\nonumber\\ &+& \left. \Theta(-m_0+\Delta) \frac{q^2}{4\pi} \epsilon^{\mu\nu\rho}A^{(-)}_{\mu}\partial_{\nu}A^{(-)}_{\rho}+\cdots \right].
\label{eacs}
\end{eqnarray}
As anticipated, this local theory is the same as the one of a $U(1)\times U(1)$ Chern-Simons theory. However, these fields are $O(2)=U(1)\rtimes \mathbb{Z}_2$ gauge fields and we shall explore the consequences in the next section. It is quite interesting to see how the pattern of phases described by Eq. (\ref{phases}) manifests here. In the trivial case, when $|m_{0}| >\Delta, ~ \text{with}~ m_0>0$, both Chern-Simons coefficients vanish, which is expected for a topologically trivial phase. In the superconducting phase, when $|m_{0}|< \Delta$, only one of the Chern-Simons terms contributes, namely, the one associated with $A^{(-)}$. Finally, in the QAH phase, for $|m_{0}|>\Delta, ~ \text{with}~ m_0<0$, both Chern-Simons are nonvanishing.

\section{Superconducting Phase and $U(1)/\mathbb{Z}_2$  Orbifold}\label{orbifold}

We shall focus on the superconducting phase, $|m_{0}|< \Delta$, whose low-energy effective theory is then described by a single $O(2)$ Chern-Simons theory at the level $q^2$,
\begin{equation}
S_{eff}[A^{(-)}]= \int d^3x\, \frac{q^2}{4\pi} \epsilon^{\mu\nu\rho}A^{(-)}_{\mu}\partial_{\nu}A^{(-)}_{\rho}.
\label{o2CS}
\end{equation}
We can in principle proceed by studying the $O(2)$ CS theory itself (see for example \cite{Amano,Fradkin,Barkeshli}). However, an insightful way to unveil the spectrum of the $O(2)$ Chern-Simons gauge theory is through its relation with rational conformal field theory (RCFT), which goes back to the classic work by Moore and Seiberg \cite{Moore}. The relationship between CS and RCFT is specially suitable for describing topological phases of matter, because of the presence of edge states described by conformal field theories whenever the system is defined on a manifold with a physical boundary. We can consider, for example, the spatial manifold as a disk $D$ with a boundary $S^1$. If the time coordinate is also a circle $S^1$, then the resulting conformal field theory will be defined on a torus $T^2=S^1\times S^1$. 
		
In the $O(2)$ case, the one-to-one correspondence between Chern-Simons and RCFT occurs when $q^2$ is of the form $q^2\equiv 2N$, with $N$ being a positive integer. We see that only for even values of the charge $q$ there is a solution with both $q$ and $N$ integers. The simplest case is $N=2$, which corresponds to $q=2$, and we shall discuss that this case describes precisely the physical properties of the superconducting phase.


\subsection{Extension of the Chiral Algebra and the Orbifold}
			
One of the key results of Moore and Seiberg \cite{Moore} is that the edge theory associated with $O(2)$ CS at the level $2N$ is the $U(1)/\mathbb{Z}_2$ orbifold theory at the level $2N$, which contains $N+7$ primary fields. To construct the orbifold theory we firstly consider the theory of a compact free scalar field $\varphi\sim \varphi +2\pi R$, where $R$ is the compactification radius (it is identified as $R=\sqrt{2N}$ through the bulk-edge correspondence), with a $U(1)$ current $j(z)=\partial_z\varphi$. When the square of the compactification radius is a rational number, some vertex operators become purely holomorphic and then can be added to the chiral algebra to produce an extended algebra, denoted by $\mathcal{A}_N$. Specifically, the maximal extension is obtained when the square of the compactification radius is of the form $\frac{R^2}{2}=\frac{p}{p'}$, where  $p$ and $p'$ are coprimes (their greatest common divisor is 1). In this case, the basic vertex operators
\begin{equation}
e^{\pm i \sqrt{2N}\varphi},~~~\text{with} ~~N=p p',
\end{equation}
become purely chiral and can be used to extend the algebra. The representations of the extended algebra are given by the remaining vertex operators that have local OPE (trivial monodromy) with all the generators of the extended algebra, namely,
\begin{equation}
e^{i \frac{n}{\sqrt{2N}}\varphi}, ~~~n\in \mathbb{Z}.
\end{equation}
However, only those operators for which $n$ belongs to the interval $n=0,1,...,2N-1$ are indeed primary fields. Integer values of $n$ outside this interval amounts to the insertion of one of the generators $e^{\pm i \sqrt{2N}\varphi}$ and consequently correspond to descendant operators. This introduces equivalence classes among states, organized in a $\mathbb{Z}_{2N}$ structure. Therefore, the number of representations is truncated and become finite, so that the extended algebra ends up having $2N$ representations. 

The next step is to project out the states that are noninvariant under $\mathbb{Z}_2$ symmetry, to obtain the orbifold $U(1)/\mathbb{Z}_2$. To this, we first note that under the action of the  $\mathbb{Z}_2$ symmetry $\varphi\rightarrow -\varphi$, the vertex operators $e^{i \frac{n}{\sqrt{2N}}\varphi}$ are mapped onto $e^{-i \frac{n}{\sqrt{2N}}\varphi}$. Then we use the $\mathbb{Z}_{2N}$ equivalence to bring it back to the interval $0,1, ...,2N-1$. In sum, the effect of the $\mathbb{Z}_2$ symmetry is $| n\rangle \rightarrow |-n + 2N\rangle$. With this in mind, we can determine the fate of the states under the action of the $\mathbb{Z}_2$ symmetry,
\begin{eqnarray}
\ket{0} &\rightarrow& \ket{0}\nonumber\\
\ket{1} &\rightarrow& \ket{-1+2N}\nonumber\\
\ket{2} &\rightarrow& \ket{-2+2N}\nonumber\\
&\vdots&\nonumber\\
\ket{N} &\rightarrow& \ket{N}\nonumber\\
&\vdots&\nonumber\\
\ket{2N-1} &\rightarrow& \ket{1}.
\end{eqnarray}
We see that the states $\ket{0}$ and $\ket{N}$ are invariant. From the remaining $2N-2$ states, we can form $(2N-2)/2$ invariant linear combinations
\begin{eqnarray}
\ket{1} &+& \ket{-1+2N}\nonumber\\
\ket{2} &+&\ket{-2+2N}\nonumber\\
&\vdots&\nonumber\\
\ket{N-1} &+&\ket{N+1}.
\end{eqnarray}
The primary operators creating such states are 
\begin{equation}
\cos\left(\frac{n}{\sqrt{2N}}\varphi\right),~~~ n=1,...,N-1.
\end{equation}
In addition, we have the operators $\mathbb{1}$ and $e^{i\sqrt{\frac{N}{2}}\varphi}$, where the latter one can be split in two independent parts 
\begin{equation}
\sin\left( \sqrt{\frac{N}{2}}\varphi\right)~~~ \text{and}~~~ \cos\left( \sqrt{\frac{N}{2}}\varphi\right).
\end{equation}

In the orbifold theory, operators that have OPE with the current that are local only after the action of the group $\mathbb{Z}_2$ are allowed. These are the so-called twist or order-disorder operators, and are defined through the following OPE with the current,
\begin{equation}
j(z)\sigma(w)\sim \frac{\tau(w)}{(z-w)^\frac12}.
\label{ope}
\end{equation}
Notice that when $z$ goes around $w$ through a $2\pi$ rotation, the OPE picks up a minus sign. In the context of the superconducting phase, the operator $ \sigma(w)$ represents a vortex located at $w$. The above relation implies the following relation between conformal dimensions of $\sigma$ and $\tau$: $h_{\tau}=h_{\sigma}+1/2$. From the OPE of the twist fields with the energy-momentum tensor $T(z)\sim j(z)j(z)$ we can compute the conformal dimensions of the twist fields: $h_{\sigma}=1/16$ and $h_{\tau}=9/16$. Twist operators always come up in pairs. In the orbifold case, we have two pairs of twist fields satisfying the OPE (\ref{ope}), namely, $\sigma_1, \tau_1$ and $\sigma_2$, $\tau_2$. They correspond to the trivial and the nontrivial representations of $\mathbb{Z}_2$ \cite{Vafa}. Therefore, we end up with 4 new operators in the theory.

Summarizing, the field content of orbifold theory, with the respective conformal dimensions, is the following:
\begin{equation}
\underbrace{\mathbb{1}}_{0},~~ \underbrace{j(z)}_{1},~~ \underbrace{\cos\left( \sqrt{\frac{N}{2}}\varphi\right)}_{\frac{N}{4}},~~ \underbrace{\sin\left( \sqrt{\frac{N}{2}}\varphi\right)}_{\frac{N}{4}},~~\underbrace{\cos\left(\frac{n}{\sqrt{2N}}\varphi\right)}_{\frac{n^2}{4N}},~~\underbrace{\sigma_a}_{\frac{1}{16}},~~\underbrace{\tau_a}_{\frac{9}{16}},
\end{equation}
with $n=1,...,N-1$ and $a=1,2$, in a total of $N+7$ fields. The respective fusion rules were derived in \cite{Vafa}. For the convenience of the reader we have included a derivation of them in the Appendix \ref{AB}, along with additional discussions of the algebra extension and the orbifold.


\subsection{$N=2$ Orbifold  $=$ Ising $\times$ Ising}
		
We are particularly interested in the values of $N$ compatible with solutions of the relation $q^2=2N$, with both $q$ and $N$ integers. The simplest case corresponds to $q=N=2$. In this case, the spectrum is quite simple and, in particular, the operators $ \cos\left( \sqrt{\frac{N}{2}}\varphi\right)$ and $\sin\left( \sqrt{\frac{N}{2}}\varphi\right)$ become fermion operators $\Upsilon_1\equiv \cos\varphi$ and $\Upsilon_2\equiv \sin\varphi$ of conformal dimension $\frac12$, with no electric charge. Therefore, they are natural candidates to be identified with the Majorana fermions appearing at the boundary of the topological  superconducting  phase. This is also compatible with the results of \cite{Vaezi}.

In addition to containing the Majorana fermions, another remarkable property of the case $N=2$, and which is crucial for its identification as the proper edge theory of the superconducting case, is that it corresponds precisely to two copies of the Ising CFT. To see this, we consider the corresponding fusion rules \cite{Vafa}:
\begin{equation}
\Phi \times \Phi = \mathbb{1} + j,~~j \times \Phi = \Phi, ~~ j\times j = \mathbb{1},~~ \Upsilon_a \times \Upsilon_a=\mathbb{1},~~\Upsilon_1 \times \Upsilon_2=j,
\label{fusion1} 
\end{equation}
\begin{equation}
\sigma_a \times \sigma_a = \mathbb{1}+\Upsilon_a,~~\sigma_1\times\sigma_2=\Phi,~~j\times \sigma_a=\tau_a,~~\Phi\times \sigma_a = \sigma_a, ~~ \Upsilon_a\times \sigma_a=\sigma_a,
\label{fusion2}
\end{equation}
where we have defined $\Phi\equiv \cos\left(\frac12\varphi\right)$. The above operator content can be decomposed into two sets,
\begin{equation}
\mathbb{1}:~(h=0)  ,~~~ \Upsilon_1:~\left(h=\frac12\right),~~~\sigma_1:~\left(h=\frac{1}{16}\right), 
\end{equation}
and 
\begin{equation}
\mathbb{1}:~(h=0)  ,~~~ \Upsilon_2:~\left(h=\frac12\right),~~~\sigma_2:~\left(h=\frac{1}{16}\right). 
\end{equation}
From the fusion rules shown in (\ref{fusion1}) and (\ref{fusion2}), we see that each one of the above sets of operators generates an independent Ising CFT with central charge $c=1/2$, namely,
\begin{equation}
\sigma_a \times \sigma_a=\mathbb{1}+\Upsilon_a,~~\Upsilon_a \times \sigma_a =\sigma_a,~~\Upsilon_a \times \Upsilon_a=\mathbb{1}.
\end{equation}
Then, taking the tensor product of the two sets of fusion rules, we recover the full set of fusion rules of the orbifold model for $N=2$. 

Recalling the discussion of Sec. \ref{Edge}, the edge theory of the superconducting phase corresponds to a single Ising CFT with chiral central charge $c=1/2$. The doubling of degrees of freedom in the orbifold theory with $N=2$ is a direct reflection of the duplication of the degrees of freedom inherent to the BdG formalism used to derive the CS effective action (\ref{eacs}), which in turn gives rise to the orbifold theory through the bulk-edge correspondence. Therefore, coming from the orbifold theory,
in order to properly account for the physical degrees of freedom, we need to halve the theory keeping thus only one of the Ising CFT's. In terms of central charge, this means that from $c=1=1/2+1/2$, we are left with only a half of the central charge, $c=1/2$, which is expected for the superconducting phase. 

After the discussion of the edge theory associated with a single $O(2)$ CS theory with level $4$, it is immediate to see that the edge theory of the QAH phase also follows from the effective action (\ref{eacs}) via bulk-edge correspondence. It emerges when $|m_{0}|>\Delta, ~ \text{with}~ m_0<0$, where both Chern-Simons theories contribute. In this case, we obtain an orbifold CFT with $N=2$ for each of the CS theories. The resulting total central charge is $c=2=1+1$, which reduces to $c=1$ after we take into account the halving mechanism for eliminating duplicated degrees of freedom.

		
\section{Conclusions}\label{conclusions}

Along this work we have studied low-energy effective theories associated with the QAH phase and with a topological superconducting phase arising from the QAH system in proximity to a pairing potential. We first derive the edge theory of the corresponding phases by transforming the 2+1 dimensional systems into a set of 1+1 dimensional quantum wires, where the edge states appear in a quite transparent way.  Next, the EFT for the QAH phase was derived directly from the microscopic model by computing the fermionic determinant at the leading order in the large gap limit. A new aspect involved in this computation is that the fermions are of nonrelativistic nature, so that the leading term in the effective action comes from several Feynman diagrams with nonrelativistic pieces that together conspire to produce a usual relativistic CS term. Of course, higher-order corrections (nontopological) are nonrelativistic. It is quite remarkable that the CS term arising from this computation does not suffer from the gauge anomaly, in contrast to the relativistic case that is plagued with a half-integer CS contribution.    

Our computation of the fermionic determinant can be used even in the presence of the superconductor pairing potential that breaks the $U(1)$ charge conservation symmetry. To this, we follow the standard treatment of superconductors where we turn to the BdG formalism and work with unconstrained spinors. In effect, this  leads to a duplication of the degrees of freedom and also introduces fictitious $U(1)$ symmetries, which can be coupled with background gauge fields. This enables us to compute the local effective action for the gauge fields in a similar way to the QAH case. The physical Hilbert space is recovered by considering the gauge fields as $O(2)$, instead of $U(1)$. In this way,  
the bulk-edge correspondence implies that the corresponding edge theory is the orbifold $U(1)/\mathbb{Z}_2$. The level of the CS, which is related to the compactification radius of the edge theory, is fixed under physical requirements of the superconducting phase. This leads to the $O(2)_4$ CS theory whose edge states are described by the $N=2$ orbifold theory, which in turn corresponds to two copies of the Ising CFT. The doubling of the degrees of freedom is a direct consequence of the way the effective theory was constructed, i.e., employing the BdG formalism and working with unconstrained spinors.

A natural extension of this work is to consider the case of Laughlin states of the fractional quantum Hall phase. While this is not described in terms of free fermions, and consequently it is very difficult to integrate out the massive fermions to produce a fermionic determinant, it can be analyzed in the framework of quantum wires \cite{Teo}. 
Recent works have shown how the effective theory containing a CS with level $m$, with $m$ being an integer odd, emerges directly from the quantum wires system through explicit identifications between quantum wires variables and gauge fields in the continuum \cite{Fontana,Imamura,Toledo}.  In this way, we see this as a promising setting to study the proximity effect in the fractional case, which in principle drive the system to a fractional topological superconducting phase.


\section{Acknowledgments}		
		
We thank Weslley Geremias dos Santos for participation in the initial stage of this project. This work is partially supported by CAPES and CNPq.

		
\appendix

\section{Coleman-Hill Theorem}\label{CH}

One important question on the one-loop generation of the CS term we described in Sec. \ref{EFT} concerns its stability. We discuss now that the topological CS term so obtained is protected against higher-order radiative corrections when the gauge field is dynamical. Actually, this result extends the Coleman-Hill theorem which asserts that in a Lorentz-invariant setting and in the absence of infrared singularities the CS term does not have corrections beyond the one-loop contribution \cite{Coleman}. To establish this result for our case, we consider the $U(1)$ conserved current following from \eqref{aqahA}. Its components are given in \eqref{0c}, which we repeat here for convenience,
\begin{equation}
J^{0}=\bar{\psi}\gamma^{0}\psi~~~\text{and}~~~J^{i} = b_{1}\bar{\psi}\gamma^{i}\psi+i b_{2}(\bar{\psi}\gamma^{i}\gamma^{j}\partial_{j}\psi-\partial_{j}\bar{\psi}\gamma^{j}\gamma^{i}\psi)+2 b_2 \bar{\psi}\psi A^i.
\end{equation}
The corresponding Ward identities following from its conservation can be derived with the help of the algebraic relation
\begin{eqnarray}
p_{0}\gamma^{0}+ b_{1}p_{i}\gamma^{i}+ b_{2} (p_{i}p^{i}+ 2k_i p^{i}+2p_i A^i)&=&(p_{0}+k_{0})\gamma^{0}+b_{1}(p_{i}+k_{i})\gamma^{i}-b_{2}(p+k)^2-m_0\nonumber\\
&-& (k_0\gamma^0 +b_1 k_i \gamma^i + b_2 k_i k^i -m_0)+ 2 b_2 p_i A^i \nonumber\\
&=& i S^{-1}(k+p) - i S^{-1}(k) + 2 b_2 p_i A^i,
\end{eqnarray}
which appears in the current vertex whenever it is contracted with the external momentum $p_{\mu}$ entering at that vertex. Indeed, by applying this identity to the set of closed fermionic loop graphs with $N>2$ amputated external gauge field lines, denoted by $\Gamma_{\mu_{1}\ldots}(p_{1},p_{2},\ldots,p_{N-1})$,  we get
\begin{equation}
p_{1}^{\mu_{1}}\Gamma_{\mu_{1}\ldots}(p_{1},p_{2},\ldots p_{N-1})=0.
\end{equation}
Taking a derivative of this expression with respect to $p_{1}^{\mu_{1}}$ and then setting $p_{1}^{\mu_{1}}=0$ leads to $\Gamma_{\mu_1\ldots}(0,p_{2},\ldots)=0$. This, in turn,  implies that $\Gamma_{\mu_{1}\ldots}(p_{1},p_{2},\ldots,p_{N-1})=O(p_1 p_2 \ldots p_{N-1})$. 

We then proceed as in \cite{Coleman} by considering  the case $N=2$. If the two trilinear gauge vertices belong to different loops then the result is $O(p^2)$ and no CS term is generated. The remaining possibility  is that the two external vertices belong to the same closed fermionic loop with some internal gauge field lines. Here, by cutting the internal  lines, we can put this graph in correspondence with another  graph without internal lines and with independent external momenta (up to the global momentum conservation). The original graph is obtained as a limit process by identifying gauge field lines and multiplying its analytical expression by the corresponding propagators.  As before, we conclude that the sum of the graphs of this type is $O(p^2)$ so that no CS term is generated. Therefore, beyond the one-loop graphs with two external lines and without  internal lines, there is no contribution to the CS term.

		
\section{Compact Boson, Chiral Algebra Extension, and Orbifold}\label{AB}

This appendix is meant to be a supplementary material to the text, covering topics that are well discussed in the CFT literature in a concise and unified way  \cite{DiFrancesco,Vafa,Ginsparg} (we follow mainly the conventions of \cite{DiFrancesco}). We start by discussing the compact boson and its action. Then, by summing over all inequivalent topological configurations we obtain the compact boson partition function, from which we read the characters of the representations. Furthermore, we extend the algebra introducing a $ \mathbb{Z}_{2N} $ structure. This process reorganizes the infinite families of the Virasoro algebra into a finite number of families in the extended algebra.


\subsubsection{Chiral Algebra Extension}
		
On the torus there are two winding directions, such that we consider the following compactification condition,
		\begin{align}
			\varphi(z+n \omega_{1}+n^{\prime} \omega_{2})=\varphi(z)+2 \pi R \left(n m+n^{\prime}m^{\prime}\right),\;\;\;\;\;\;\;\;\;\; n,n^{\prime},m,m^{\prime} \in \mathbb{Z}, \label{a10}
		\end{align}
where $ R $ and $ \omega_{i} $ are the compactification radius and directions of the torus in complex coordinates. The indices $ m $ and $ m^{\prime} $ specify inequivalent topological classes of configurations. The boson integration may be done by separating the boson field into a topologically nontrivial part $ \varphi_{m,m^{\prime}} $ and a periodic part $\tilde{\varphi}$, i.e., $\varphi=\varphi_{m,m^{\prime}}+\tilde{\varphi}$, where
\begin{equation}
\varphi_{m,m^{\prime}}=2 \pi R \left[\frac{z}{\omega_{1}} \frac{m \bar{\tau}-m^{\prime}}{\bar{\tau}-\tau}-\frac{\bar{z}}{\omega_{1}^{*}} \frac{m \tau-m^{\prime}}{\bar{\tau}-\tau}\right]
\label{a11}
\end{equation}
is compatible with \eqref{a10} and $ \tau\equiv \omega_{2}/\omega_{1} $ is the modular parameter. Since $ \partial_z \partial_{\bar{z}}\varphi_{m,m^{\prime}}=0 $, the action decomposes into $ S[\varphi]=S[\varphi_{m,m^{\prime}}]+S[\tilde{\varphi}]  $ and we can write the partition function for each topological class as 
\begin{align}
Z_{m,m^{\prime}}(\tau)=Z_{per}e^{-S[\varphi_{m,m^{\prime}}]}, 
\label{a14}
\end{align}
where $ Z_{per} $ is the result of the periodic boson integration and the remaining action is given by
\begin{align}
S[\varphi_{m,m^{\prime}}]&=\frac{1}{2\pi}\int dz d\bar{z} \partial_z \varphi_{m,m^{\prime}}\partial_{\bar{z}}\varphi_{m,m^{\prime}}\nonumber\\
&=\pi R^{2}\frac{\abs{m \tau-m^{\prime}}^{2}}{2 \text{Im}\tau}. \label{a12}
\end{align} 

Under the action of the generators of modular invariance,
\begin{align}
	\mathcal{T}: \tau \rightarrow \tau+1 \qquad \qquad \text{and} \qquad \qquad \mathcal{S}: \tau\rightarrow -\frac{1}{\tau},\label{a13}
\end{align}
the partition functions \eqref{a14} transform among themselves:
\begin{align}
		Z_{m,m^{\prime}}(\tau+1)&=Z_{m,m^{\prime}-m}(\tau),\\
	Z_{m,m^{\prime}}(-1/\tau)&=Z_{-m^{\prime},m}.
\end{align}
As this only amounts to a redefinition of the indices, the complete partition function,
\begin{align}
Z=Z_{per}\sum\limits_{m,m^{\prime}\in \mathbb{Z}}e^{-\pi R^{2}\frac{\abs{m \tau-m^{\prime}}^{2}}{2 \text{Im}\tau}},
\end{align}
is modular invariant.
		
In order to identify the Virasoro characters from the partition function, it is useful to recast it in a manner that better reflects its holomorphic separation. To this end, we apply the Poisson resummation formula, which states that for two sums over the integers
\begin{align}
\sum\limits_{m^{\prime} \in \mathbb{Z}}f(m^{\prime})=\sum\limits_{n\in \mathbb{Z}}\tilde{f}_{n}, \label{a18}
\end{align}
where $ \tilde{f}_{k} $ is the Fourier transform of $ f(x) $,
\begin{align}
\tilde{f}_{n}=\int\limits_{-\infty}^{\infty}\dd{x} f(x) e^{-2 \pi i x n}. \label{a19}
\end{align}
	
Applied to the compact boson partition function, the resummation formula yields the familiar expression for the compact boson partition function
\begin{align}
Z= \sum_{n, m \in \mathbb{Z}} \chi_{n,m}(q) \bar{\chi}_{n,m}(\bar{q}), \label{a20}
\end{align}
where the Virasoro characters
\begin{equation}
\chi_{n,m}(q)=\frac{1}{\eta(q)}q^{(n / R+m R / 2)^{2} / 2}~~~\text{and}~~~ \bar{\chi}_{n,m}(\bar{q})=\frac{1}{\bar{\eta}(\bar{q})}\bar{q}^{(n / R-m R / 2)^{2} / 2},
\label{a21}
\end{equation}
with $q\equiv e^{2\pi i \tau}$ and $\bar{q}\equiv e^{-2 \pi i \bar{\tau}}$, are associated with the conformal families of the theory. Each family contains an infinite number of conformal fields, which are generated by successive application of the positive modes of the energy-momentum tensor on the primary field of the family. As expected for a free scalar CFT, the primary fields are the vertex operators 
		\begin{align}
			V_{n,m}=e^{i \left(n/R+mR/2\right)\varphi} \qquad \text{and}\qquad \bar{V}_{n,m}=e^{i \left(n/R-mR/2\right)\bar\varphi},
		\end{align}
		with conformal dimension
		\begin{align}
			h_{n, m}=\frac{1}{2}(n/ R+m R / 2)^{2} \qquad\text{and} \qquad \bar{h}_{n, m}=\frac{1}{2}(n / R-m R / 2)^{2}. \label{a22}
		\end{align}
At this point we can also infer the invariance of the partition function under $ R \rightarrow 2/R $, as it amounts to the exchange $ n \leftrightarrow m $.
		
So far the partition function in Eq. \eqref{a20} embodies only the Virasoro algebra. In order to add the $\mathbb{Z}_{2N} $ algebra, we need to perform an extension of the Virasoro algebra. To do so, we restrict the compactification radius according to
		\begin{align}
			\frac{R^{2}}{2}=\frac{p}{p^{\prime}} \label{a23}
		\end{align}
with $ p $, $ p^{\prime} $ natural coprimes and introduce the new indices
		\begin{align}
			\begin{array}{lll}
				n=2 p n^{\prime}+r, ~~~ & 0 \leq r \leq 2 p-1, & ~~~ r,n^{\prime} \in \mathbb{Z}; \\
				m=2 p^{\prime} m^{\prime}+s,~~~ & 0 \leq s \leq 2 p^{\prime}-1, & ~~~ s,m^{\prime} \in \mathbb{Z}; \label{a24}
			\end{array}
		\end{align}
such that now we sum over the integers
		\begin{align}
			\begin{array}{lll}
				u=n^{\prime}+m^{\prime},&~~~u \in \mathbb{Z};\\
				l=p^{\prime}r+p s,&~~~ l\in \mathbb{Z}. \label{a25}
			\end{array}	
		\end{align}
In terms of the new indices the partition function reads
		\begin{align}
			Z=\sum\limits_{l,\bar{l}} \chi_{l}(q)\bar{\chi}_{\bar{l}}(\bar{q}), \label{a26}
		\end{align}
where the limits of the sum over $ l $ are yet to be specified and the extended algebra characters are given by
		\begin{align}
			\chi_{l}(q)=\frac{1}{\eta(q)}\sum\limits_{u \in \mathbb{Z}}q^{N\left(u+l/2N\right)^{2}}, \label{a27}
		\end{align}
where $ N\equiv p p^{\prime} $. This is the maximal extension of the algebra.
				
From the definition of the extended character \eqref{a27} it is easy to derive that $ \chi_{l}=\chi_{l+2N} $. As we associate each character to a primary field, there are $ 2N$ primary vertex fields, given by
\begin{align}
V_{l}= e^{i l \varphi/\sqrt{2N}}. 
\label{a27.1}
\end{align}
Consequently, two vertex fields, $ V_{l} $ and $ V_{l+2N} $, must necessarily belong to the same Verma module and there must exist operators which connects them. This is analogous to the role played by the $ L_{-n} $ in the Virasoro algebra. We find the ladder operators to be
\begin{align}
\Gamma_{\pm}=e^{\pm i \sqrt{2N}\varphi}, 
\label{a30}
\end{align}
with conformal dimensions $h_{\Gamma_{\pm}}=N$.

%

It might be useful to extend the notion of a primary field of the extended algebra $ \mathcal{A} $ by requiring that it must be annihilated by all the positive modes of the currents that generate the corresponding algebra. In terms of OPE, this is equivalent to defining a primary field of $ \mathcal{A} $ to have the OPE
\begin{align}
\mathcal{J}(z) \Phi(0)=z^{-h_{\mathcal{J}}}\Phi(0)+\text{less singular}, \label{a29}
\end{align}
where $\mathcal{J}$ is an algebra generating current. This equation should be understood in the sense that the OPE of $ \mathcal{J} $ with a primary field has a \textit{maximum} allowed singularity; i.e., less singular OPEs are allowed. This is nothing else than the Virasoro primary field condition generalized to an extended algebra. In this more familiar case, the generators of the algebra are the holomorphic and antiholomorphic parts of the energy momentum tensor, $ T(z)\equiv T_{ z z }(z) $  and $ \bar{T}(\bar{z})\equiv  T_{\bar{z}\bar{z}} (\bar z)$, with conformal dimensions $ h=2 $ and $ \bar{h} =2$, respectively.
		
The maximal extension is equivalent to choosing the $ U(1) $ current $j$, the ladder operators $ \Gamma_{\pm} $, and the energy-momentum tensor $ T=T_{z,z}(z) $ to be the set of algebra generating currents $ \mathcal{J} $. Applying the primary field condition \eqref{a29} to the ladder operators, it follows that
\begin{align}
\Gamma_{\pm}(z)V_{l}(0)&=e^{\pm i \sqrt{2N} \varphi(z)}e^{i l \varphi(0) /\sqrt{2N}}=e^{i \left(l\pm 2N\right)\varphi(0)/\sqrt{2N}}e^{\mp l \expval{\phi(z)\phi(0)}} \nonumber\\
&=z^{\pm l}V_{l\pm 2N}(0). \label{a31}
\end{align}
Comparing with the primary field OPE, we see that the spectrum of primary fields is given by the $ V_{l} $ with $ -(N-1)\leq l\leq N $ and the $ U(1) $ partition function reads
\begin{align}
Z_{U(1)}=\sum\limits_{l,\bar{l}=-(N-1)}^{N} \chi_{l}(q)\bar{\chi}_{\bar l}(\bar{q}) \label{a32}.
\end{align}
It is standard in the literature to choose the range of $ l $ in the definition of the primaries, Eq. \eqref{a27.1}, to be $ l=0,1,2, \cdots,2N-1 $. We use this convention in the main body of this paper.


\subsubsection{Orbifold}

Now that we have considered the extension of the chiral algebra we are in a position to study the  $U(1)/\mathbb{Z}_{2} $ orbifold. Such a theory is obtained by considering only field configurations of $ U(1) $ that are invariant under the $\mathbb{Z}_{2}$ group action, $ \varphi \rightarrow - \varphi $. To do so, we extend the scalar field compactification condition such that it admits twists when going around the torus
		\begin{align}
			\varphi\left(z+n \omega_{1}+n^{\prime}\omega_{2}\right)=e^{2 \pi i(n v+n^{\prime} u)} \varphi(z), \label{a36}
		\end{align}
where $v,u = 0, \frac{1}{2}$ for periodic and antiperiodic boundary conditions, respectively. At this point it is convenient to identify the windings to be oriented along one of the Cartesian directions. We choose $ \omega_1$ to represent winding around the space direction and $\omega_2$ around time. 
		
To each pair $v,u$, we associate a partition function $ Z_{v,u} $. The complete orbifold partition function is given by the sum of the $ Z_{v,u} $. The untwisted sector, $ Z_{0,0} $, corresponds to  the $ Z_{U(1)}$ of the previous discussion, so that we are left to calculate the partition function for the twisted sectors. For the antiperiodic condition along at least one of the torus directions, we separate the partition function into its holomorphic blocks
\begin{align}
Z_{vu}=\abs{f_{vu}}^{2},  
\label{a37}
\end{align}
where the conformal blocks are given by the character
\begin{align}
f_{vu}\equiv\Tr q^{L_{0}^{(v)}-1/24}.
\end{align}
The superscript on $ L_{0}^{(v)} $ means that $ L_{0} $ should be taken with the corresponding boundary conditions along the space direction, specified by $ v$.
		
The computation of the trace in (\ref{a37}) requires some care in the case of antiperiodic boundary conditions along the time, i.e., when $u=\frac12$. To understand this, let us examine a correlation function with the time antiperiodic boundary condition
\begin{align}
\expval{T\varphi(z)X(\lbrace z_{i}\rbrace)}, 
\end{align}
where $ T $ is the time ordering and $ X(\lbrace z_{i}\rbrace) $ stands for the insertion of any number of boson operators at the positions $ z_{i}$. Now, consider that we take $ \varphi $ along a continuous path from  $ z $ to $ z+\omega_{2} $. Because of the time ordering $ \varphi $ will pass over all the insertions on $ X $ in succession and return to the starting place. As we are dealing with bosons this operation does not pick a sign. On the other hand, under this process $ \varphi $ picks up a sign when $u=1/2$. To reconcile this discrepancy, we can introduce an operator $ \mathcal{G} $ on every correlation function of $ \varphi $ in the case of the antiperiodic boundary condition along time, where $ \mathcal{G} $ is the operator that implements the $ \mathbb{Z}_{2} $ symmetry, $ \mathcal{G}\varphi\mathcal{G}^{-1}=-\varphi $ \footnote{This discussion is similar to what happens on compact fermions, where in the time periodic case we introduce $ \mathcal{G}=(-1)^{F} $, with $ F $ being the fermion number operator.}. In particular, the traces with $u=\frac12$ in the partition function \eqref{a37} must also include the operator $\mathcal{G}$.
		
To illustrate the procedure, we consider the specific case of $ f_{0,1/2} $. The above prescription leads us to write the holomorphic blocks of the partition function as
\begin{align}
f_{0,1/2}=\Tr \mathcal{G}q^{L_{0}^{(0)}-1/24}=q^{-1/24}\Pi_{n>0}\tr^{(n)} \mathcal{G}_{n}q^{a_{-n}a_{n}},
\end{align}
where the trace on the right-hand side is defined to be taken over states with fixed $ n $. That is,
\begin{align}
	\tr^{(n)}\mathcal{G}_{n}q^{a_{-n}a_{n}}&=\sum_{k=0}^{\infty}\bra{k}\mathcal{G}_{n}q^{a_{-n}a_{n}}\ket{k}=\sum_{k=0}^{\infty}\bra{0}\left(a_{n}\right)^{k} \mathcal{G}_{n}q^{a_{-n}a_{n}} \left(a_{-n}\right)^{k}\ket{0}\nonumber\\
	&=\sum_{k,m=0}^{\infty}\frac{\left(2\pi i \tau\right)^{m}}{m!} \bra{k}\mathcal{G}_{n}\left(a_{-n}a_{n}\right)^{m} \left(a_{-n}\right)^{k}\ket{0} \nonumber\\
	&=\sum_{k,m=0}^{\infty} \frac{\left(2\pi i \tau\right)^{m}}{m!} \bra{k} \mathcal{G}_{n} \left(a_{-n}a_{n}\right)^{m-1}\left(nk\right)\left(a_{-n}\right)^{k}\ket{0}\nonumber\\
	&=\sum_{k,m=0}^{\infty} \frac{\left(2\pi i \tau n k\right)^{m}}{m!}\bra{k}\mathcal{G}_{n} \ket{k}=\sum_{k=0}^{\infty}(-1)^{k} q^{k n} \nonumber\\
	&= \frac{1}{1+q^{n}},
\end{align}
where we have used the commutation relation $ \left[a_{-n}a_{n},\left(a_{-n}\right)^{k}\right]=n k \left(a_{-n}\right)^{k} $ to get from the second to the third line. Returning to the conformal block, we obtain
		\begin{align}
			f_{0,1/2}=q^{-1/24}\prod\limits_{n=1}^{\infty}\frac{1}{1+q^{n}}=\frac{\sqrt{\theta_{3}\theta_{4}}}{\eta}\approx\frac{1}{\eta}\left(1-2q+2q^{4}-2q^{9}+2q^{16}+\cdots\right),
		\end{align}
where $ \theta_{i} $ are the Jacobi theta functions, as defined in \cite{DiFrancesco}.

We calculate the next two conformal blocks in a similar way,
		\begin{align}
			f_{1/2,0}&=q^{1/48}\prod\limits_{n \in \mathbb{N}+1/2}\Tr q^{a_{-n}a_{n}}=q^{1/48}\prod\limits_{n \in \mathbb{N}+1/2} \sum\limits_{N=0}^{\infty}q^{nN}=q^{1/48}\prod\limits_{n \in \mathbb{N}+1/2}\frac{1}{1-q^{n}}\nonumber\\
			&=\frac{\sqrt{\theta_{2}\theta_{3}/2}}{\eta}\approx\frac{q^{1/16}}{\eta}\left(1+q^{1/2}+q^{3/2}+q^{3}+q^{5}+q^{15/2}+\cdots\right),
\end{align}			
and for the next one,
\begin{align}
			f_{1/2,1/2}&=q^{1/48}\prod\limits_{n \in \mathbb{N}+1/2}\Tr\mathcal{G}_{n} q^{a_{-n}a_{n}}=q^{1/48}\prod\limits_{n \in \mathbb{N}+1/2} \sum\limits_{N=0}^{\infty}(-1)^{N}q^{nN}=q^{1/48}\prod\limits_{n \in \mathbb{N}+1/2}\frac{1}{1+q^{n}}\nonumber\\
			&=\frac{\sqrt{\theta_{2}\theta_{4}/2}}{\eta}\approx\frac{q^{1/16}}{\eta}\left(1-q^{1/2}-q^{3/2}+q^{3}+q^{5}-q^{15/2}+\cdots\right).
		\end{align}

The orbifold partition function then reads 
		\begin{align}
			Z_{orb}=\frac{1}{2}\left[Z_{U(1)}+\abs{f_{0,1/2}}^{2}+\abs{f_{1/2,0}}^{2}+\abs{f_{1/2,1/2}}^{2}\right].
		\end{align}
It is important to note that these conformal blocks individually are not invariant under the modular transformations \eqref{a13}. Under $ \mathcal{T} $ the conformal blocks $ f_{v,u} $ transform as
\begin{align}
	f_{0,1/2}(\tau+1)&=e^{-i \pi/12}f_{0,1/2}(\tau)\nonumber\\
	f_{1/2,0}(\tau+1)&=e^{i \pi/24}f_{1/2,1/2}(\tau)\nonumber\\
	f_{1/2,1/2}(\tau+1)&=e^{i \pi/24}f_{1/2,0}(\tau)
\end{align}
 and under $ \mathcal{S} $,  $ f_{v,u}(-1/\tau)=f_{u,v}(\tau) $. In this way, even though the conformal blocks themselves are not modular invariant, the total partition function is.

Notice that $ f_{1/2,0} $ and $ f_{1/2,1/2} $ mix under $ \mathcal{T} $ and a similar thing happens with $ f_{0,1/2} $ and $ \chi_{0} $  of the $ U(1) $ partition function as given in \eqref{a27}. Furthermore, these are the only terms that can be written in the form $ \abs{f}^{2} $. Let us check their contribution to the partition function:
\begin{align}
			Z_{orb} &\supset \frac{1}{2}\left(\chi_{0}\bar{\chi}_{0}+\abs{f_{0,1/2}}^{2}+\abs{f_{1/2,0}}^{2}+\abs{f_{1/2,1/2}}^{2}\right)\nonumber\\
			&=\frac{1}{4}\left[\left(\chi_{0}+f_{0,1/2}\right)\left(\bar{\chi}_{0}+\bar{f}_{0,1/2}\right)+\left(\chi_{0}-f_{0,1/2}\right)\left(\bar{\chi}_{0}-\bar{f}_{0,1/2}\right)+\right.\nonumber\\
			&+\left.\left(f_{1/2,0}+f_{1/2,1/2}\right)\left(\bar{f}_{1/2,0}+\bar{f}_{1/2,1/2}\right)+\left(f_{1/2,0}-f_{1/2,1/2}\right)\left(\bar{f}_{1/2,0}-\bar{f}_{1/2,1/2}\right)\right].
		\end{align}
This motivates us to define the orbifold characters and their associated primary fields as
		\begin{align}
			\begin{array}{lll}
			\mathbb{1}&: \;\;\chi^{orb}_{\mathbb{1}}\equiv\left(\chi_{0}+f_{0,1/2}\right)/2\approx \eta^{-1}\left(1-q+q^2+q^4+\cdots\right), &h_{\mathbb{1}}=0;\\
			j&:\;\; \chi_{j}^{orb}\equiv\left(\chi_{0}-f_{0,1/2}\right)/2\approx \eta^{-1}q\left(1+q-q^{3}+q^{7}+\cdots\right), &h_{j}=1; \\
			\sigma&:\;\; \chi_{\sigma}^{orb}\equiv\left(f_{1/2,0}+f_{1/2,0}\right)/2\approx \eta^{-1}q^{\frac{1}{16}}\left(1+q^{3}+q^{5}+q^{14}+\cdots\right), \qquad &h_{\sigma}=\frac{1}{16};\\
			\tau&:\;\; \chi_{\tau}^{orb}\equiv\left(f_{1/2,0}-f_{1/2,0}\right)/2\approx \eta^{-1}q^{\frac{9}{16}}\left(1+q+q^{7}+q^{10}+\cdots\right),&h_{\tau}=\frac{9}{16};
			\end{array}
		\end{align}
where the expansions are given for $ N=2 $ and the antiholomorphic sector characters are defined similarly. With the new characters, this contribution to the partition function can be rewritten as
		\begin{align}
			Z_{orb}\supset \abs{\chi_{\mathbb{1}}^{orb}}^{2}+\abs{\chi_{j}^{orb}}^{2}+\abs{\chi_{\sigma}^{orb}}^{2}+\abs{\chi_{\tau}^{orb}}^{2}.
		\end{align}
	
The remaining terms in the orbifold partition function are the $ U(1) $ characters, $ \chi_{k\neq0} $. Let us consider 
		\begin{align}
			Z_{orb}\supset\frac{1}{2}\sum_{k=-(N-1)}^{N-1 }\!\!\!\!\!\!\!{}^{\prime}\;\;\;\;\sum_{\bar{k}=-(N-1)}^{N-1} \!\!\!\!\!\!\!{}^{\prime}\chi_{k}\bar{\chi}_{\bar{k}}=2 \sum \limits_{k,\bar{k}=1}^{N-1}\chi_{k}\bar{\chi}_{\bar{k}}\equiv 2 \sum \limits_{k,\bar{k}=1}^{N-1}\chi_{k}^{orb}\bar{\chi}^{orb}_{\bar{k}},
		\end{align}
where the primes on the sums mean that the $ k,\bar{k}=0 $ are excluded. We associate with these characters the $ N-1 $ primary fields
		\begin{align}
			\phi_{k}=\cos \left(\frac{k}{\sqrt{2N}}\varphi\right),\qquad  h_{k}=\frac{k^{2}}{4N} \qquad\text{for} \qquad k=1,2,\cdots,N-1.
		\end{align}
For the remaining $ k=N $ character, we define
		\begin{align}
			\phi^{i}_{N}:\;\;	\chi^{orb}_{N}\equiv \frac{1}{2}\chi_{N}\approx \eta^{-1}q^{1/2}\left(1+q^{4}+q^{12}+q^{24}+\cdots\right), \qquad  h_{\phi_{i}^{N}}=\frac{N}{4},
		\end{align}
		where the expansion is given for $ N=2 $. Its contribution to the partition function reads
		\begin{align}
			Z_{orb}\supset 2 \chi_{N}^{orb}\bar{\chi}_{N}^{orb}.\;\;\;\;\;\;\;\;\text{(No index sum)}.
		\end{align}
Thus, the $ U(1)/Z_{2} $ orbifold partition function reads
\begin{align}
	Z_{orb}=\abs{\chi_{\mathbb{1}}^{orb}}^{2}+\abs{\chi_{j}^{orb}}^{2}+\abs{\chi_{\sigma}^{orb}}^{2}+\abs{\chi_{\tau}^{orb}}^{2}+2\sum\limits_{k,\bar{k}=1}^{N} \chi_{k}^{orb}\bar{\chi}_{k}^{orb}.
\end{align}

Our final task is to determine the fusion rules of the model. One way to do so is from the Verlinde formula
\begin{align}
	N_{i,j}^{k}=\sum\limits_{n}\frac{S_{i,n}S_{j,n}S_{k,n}}{S_{\mathbb{1},n}}, \label{a50}
\end{align}
where $ N_{i,j}^{k} $ is the fusion coefficients \footnote{In general the fusion matrices as defined in \eqref{a50} and \eqref{a51} are not precisely the same, but are related by a raising and lowering matrix, which in the present case is diagonal \cite{DiFrancesco}.}
\begin{align}
	\phi_{i}\times \phi_{j}=\sum\limits_{k} N_{i,j}^{k}\phi_{k} \label{a51}
\end{align}
		and $ S_{i,j} $ determines how the characters behave under the modular transformation $ \mathcal{S} $, namely,
		\begin{align}
			\chi_{i}^{\prime}=\chi_{i}(-\frac{1}{\tau})=\sum\limits_{j}S_{i,j}\chi_{j}(\tau).
		\end{align}

We will consider the case of even $ N $, which is relevant for the discussion of the Sec.  \ref{orbifold} (the case of odd $ N $ can be similarly constructed \cite{Vafa}). Let us discuss the transformation of some characters starting with the identity
		\begin{align}
			\chi_{\mathbb{1}}^{\prime}=\frac{1}{\sqrt{8N}}\left[\chi_{\mathbb{1}}+\chi_{j}+2 \sum\limits_{k=1}^{N-1}\chi_{k}+\chi_{N}^{i}+2\sqrt{N}\chi_{\sigma}+2\sqrt{N}\chi_{\tau}\right].
		\end{align}
The factor of two for the $ \sigma $ and $ \tau $ representations should be understood as a reflection of the fact that there are two representations for $ \sigma $ and $ \tau $. Therefore, when building the $ S $ matrix, the contribution for the twists should read $ \sqrt{N}\left[\chi_{\sigma^{1}} +\chi_{\sigma^{2}} +\chi_{\tau^{1}} +\chi_{\tau^{2}}\right] $, where the symmetry of the splitting between $ \sigma^{1} $ and $ \sigma^{2} $ is a consequence of the unitarity of the $ S $ matrix. 

Let us proceed to the next one. A naive analysis of the modular transformation leads us to 
\begin{align}
			\chi_{N}^{i \prime}=\frac{1}{\sqrt{8N}} \left[\chi_{\mathbb{1}}+\chi_{j}+2\sum\limits_{k=1}^{N-1}(-1)^{k}\chi_{k}+(-1)^{N}\chi_{N}\right],
\end{align}
but this is still not correct. This modular transformation, as is presented, hides the twist characters. In fact, it only tells us that the sum of the contributions for the characters should cancel out, similarly to the last calculation. In this manner, the contributions for $ \chi_{\sigma_{1}} $ should cancel out the contribution for $ \chi_{\sigma_{2}} $, and similarly for $ \chi_{\tau_{i}}$. We implement this by adding to the brackets above the term $ x \sigma_{i,j} \left(\chi_{\sigma_{j}}+\chi_{\tau_{j}}\right)$, where $ x $ is determined by demanding that the $ S $ matrix must be unitary and $ \sigma_{i,j}\equiv 2 \delta_{i,j}-1 $. Similar considerations are needed for the transformation of $ \chi_{k} $, $ \chi_{\sigma} $ and $ \chi_{\tau} $, after which we obtain the $ S $ matrix for the orbifold
\begin{center}
	\begin{tabular}{c|c c c c c c}		
				& $\mathbb{1}$ & $ j $ & $ \phi^{i}_{N} $ & $ \phi_{k} $ &$ \sigma_{i} $  &$ \tau_{i} $   \\
				\hline
				$ \mathbb{1} $	& 1 & 1 & 1 & 2 & $ \sqrt{N} $ & $ \sqrt{N} $ \\
				
				$ j $	& 1 & 1 &1  & 2 & $ -\sqrt{N} $ & $ -\sqrt{N} $  \\
				
				$ \phi_{N}^{j} $	& 1 & 1 & 1 & $ 2(-1)^{k} $ & $ \sigma_{i,j}\sqrt{N} $ & $ \sigma_{i,j}\sqrt{N} $ \\
				
				$ \phi_{k^{\prime}} $	& 2 & 2 & $ 2(-1)^{k^{\prime}} $ & $ 4\cos(\frac{\pi k k^{\prime}}{N}) $ & 0 & 0 \\
				
				$ \sigma_{j} $	& $ \sqrt{N} $ & $ -\sqrt{N} $ & $ \sigma_{i,j}\sqrt{N} $ & 0 & $ \sqrt{2N} \delta_{i,j}$ 	 & $ -\sqrt{2N}\delta_{i,j} $ \\
				
				$ \tau_{j} $	& $ \sqrt{N} $ & $ -\sqrt{N} $ & $ \sigma_{i,j}\sqrt{N} $ & 0 & $ -\sqrt{2N}\delta_{i,j} $ & $ \sqrt{2N}\delta_{i,j} $ \\
	\end{tabular},
\end{center}
where we have factored out $ \left(8N\right)^{-1/2} $.
		
With the $ S $ matrix at hand, we only need to run the Verlinde formula to find the fusion rules:
\begin{align}
	j \times j=1, ~~~~\phi^{i}_{N}\times \phi_{N}^{i}=1,~~~~\phi_{k}\times\phi_{k^{\prime}}=\phi_{k+k^{\prime}}+\phi_{k-k^{\prime}}, \nonumber\\
	j\times\phi_{k}=\phi_{k},~~~~j \times \sigma_{i}=\tau_{i} ~~~\text{and}~~~\phi_{k}\times \phi_{k}=1+j+\phi_{2k}.
\end{align}


\end{document}